\documentclass[fleqn,usenatbib]{mnras}

\usepackage{newtxtext,newtxmath}

\usepackage[T1]{fontenc}

\DeclareRobustCommand{\VAN}[3]{#2}
\let\VANthebibliography\thebibliography
\def\thebibliography{\DeclareRobustCommand{\VAN}[3]{##3}\VANthebibliography}

\usepackage[dvipdfmx]{graphicx}	
\usepackage{amsmath}	

\newcommand{\mr}[1]{\mathrm{#1}}
\newcommand{\diff}[2]{\frac{\mr{d}{#1}}{\mr{d}{#2}}}

\newcommand{\da}{\mr{d}a}
\newcommand{\dt}{\mr{d}t}
\newcommand{\dm}{\mr{d}m}
\newcommand{\dmd}{\mr{d}m_\mr{d}}
\newcommand{\amax}{a_\mr{max}}
\newcommand{\amin}{a_\mr{min}}
\newcommand{\rhox}{\rho_\mr{X}}
\newcommand{\fx}{f_\mr{X}}
\newcommand{\md}{m_\mr{d}}

\newcommand{\intamm}{\int^{\amax}_{\amin}}

\newcommand{\Pegase}{\textsc{P\'egase}}

\title[New SED model with dust evolution]{A new galaxy spectral energy distribution model consistent with the evolution of dust}

\author[K. Y. Nishida et al.]{
Kazuki Y. Nishida,$^{1}$\thanks{E-mail: nishida.kazuki@nagoya-u.jp}
Tsutomu T. Takeuchi,$^{1,2}$
Takuma Nagata,$^{1}$
and Ryosuke S. Asano$^{1}$
\\
$^{1}$Division of Particle and Astrophysical Science, Nagoya University, Furo-cho, Chikusa-ku, Nagoya, 464--8602, Japan\\
$^{2}$The Research Center for Statistical Machine Learning, The Institute of Statistical Mathematics, 10-3 Midori-cho, Tachikawa, Tokyo 190--8562, Japan
}

\date{Accepted XXX. Received YYY; in original form ZZZ}

\pubyear{2020}

\begin{document}
\label{firstpage}
\pagerange{\pageref{firstpage}--\pageref{lastpage}}
\maketitle

\begin{abstract}
The spectral energy distribution (SED) of galaxies provides fundamental information on the related physical processes. 
However, the SED is significantly affected by dust in its interstellar medium.
Dust is mainly produced by asymptotic giant branch stars and Type II supernovae.
In addition, the dust mass increases through the metal accretion, and the grain size changes by the collisions between the grains.
The contribution of each process and the extinction depend on the size distribution.
Therefore, the SED model should treat the evolution of the dust mass and size distribution.
In spite of the importance of dust evolution, many previous SED models have not considered the evolution of the total mass and size distribution in a physically consistent manner.
In this work, we constructed a new radiative transfer SED model, based on our dust evolution model consistent with the chemical evolution. 
To reduce the computational cost, we adopted the mega-grain and the one-dimensional plane parallel galaxy approximation.
As a fiducial case, we calculated Milky Way-like galaxy SEDs at various ages under the closed-box model.
We found that a galaxy at the age of 100~Myr does not produce small grains such as polycyclic aromatic hydrocarbons.
After 1~Gyr, we observed a drastic increase of infrared emission and attenuation caused by a rapid increase of dust mass.
This phenomenon can be treated appropriately for the first time by our new model.
This model can be used for the SED fitting to a galaxy at any stage of evolution.
\end{abstract}

\begin{keywords}
dust, extinction -- galaxies: evolution -- ISM: evolution -- radiative transfer -- galaxies: ISM -- galaxies: disc
\end{keywords}



\section{Introduction}
The spectral energy distribution (SED) fitting is a fundamental method to extract the information of the physical processes in galaxies (e.g., star formation rate: SFR, stellar mass, dust mass) from observational data. 
Stars emit photons with wavelengths ranging from ultraviolet (UV) to near-infrared (NIR).
Dust grains absorb and scatter the photons emitted from stars, and re-emit the absorbed energy at mid-infrared (MIR) to far-infrared (FIR).
In addition to the radiative aspect of a galaxy, dust grains promote the formation of hydrogen molecules on the surface of the grains \citep[e.g.,][]{Hollenbach1979, Hirashita2002, Cazaux2005}.
Since hydrogen molecules are one of the fundamental ingredients of the star formation, dust grains directly activate the star formation in galaxies. 

Dust is a solid grain consisting of the elements heavier than helium, and is produced by stellar mass loss and supernovae (SNe).
Outflows from low- and intermediate-mass stars during the thermally pulsing asymptotic giants branch (TP-AGB) phase and Type II SNe (SNe II) are considered to be the primary sources of dust \citep[e.g.,][]{Nozawa2007, Bianchi2007, Zhukovska2008}.
Dust grains are formed by condensation of heavy elements in the atmosphere of massive stars, and only the grains that could survive the reverse shock of the SN are finally expelled into the interstellar medium (ISM).
Then, the blast waves from SNe propagating in the ISM also destroy dust grains \citep[e.g.,][]{Jones1994, Jones1996, Nozawa2003, Nozawa2006, Zhukovska2008, Yamasawa2011}.
This destruction process has been confirmed by observations of several supernova remnants \citep[e.g.,][]{Borkowski2006, Arendt2010}.
Details of the dust grain survival still remain controversial.
It might depend on their composition and size\footnote{Terminologies such as grain size, size distribution, etc. are often used in articles related to dust. 
Throughout this paper, when we mention ``size'' of dust grain, it always mean the dust grain radius under the assumption of spherical shape.} \citep[e.g.,][]{Nozawa2007, Gall2014, Slavin2020}, and others claim that it depends on the clumpiness of the ejecta \citep[e.g.,][]{Biscaro2016}.
In addition, \citet{Matsuura2019} argue that the dust destruction by the SN are suppressed because the atoms can stick to the surviving dust grain in the passage of the forward shock region and it can reform or increase dust grains.
Observations of SN remnants (SNRs) do not give a final answer, since dust mass and composition are significantly different among observed SNRs.

In addition to the dust production from stars, dust growth in the ISM is also an important process, and necessary to explain the large amount of dust present in galaxies \citep[e.g.,][]{Asano2013a, Zhukovska2014, Michaowski2015, Lesniewska2019}.
In the cold phase of ISM, metal is accreted onto the dust grain surface to increased the size and total mass of the dust \citep[e.g.,][]{Dwek1998, Zhukovska2008, Michaowski2010a, Hirashita2011, Asano2013a}.
The size distribution of the dust is also changed through the collisions between dust grains \citep[e.g.,][]{Yan2004, Jones1996, Hirashita2009, Kuo2012}.

Which process controls the mass of dust varies greatly, depending on the age and the environment of the galaxy, and it is still actively debated from several different points of view.
The SNe II dominates the dust production, especially in very young galaxies, because SNe II have a shorter lifetime ($<30$~Myr) than AGB stars ($>150$~Mry) \citep[e.g.,][]{Morgan2003, Marchenko2006, Dwek2007, Valiante2009, Gall2011a, Gall2011b, Liu2019, DeLooze2020, Burgarella2020, Nanni2020}.
We should note, however, that the contribution of AGB cannot be ignored even in galaxies with the age of 500~Myr, when the star formation rate (SFR) is high \citep{Valiante2009}.

The debate of dust grain growth in high-$z$ galaxies has not been settled.
In high-$z$ galaxies, several studies claim that the process is not effective because there is not enough time to growth, low gas density, and high temperature \citep[e.g.,][]{Ferrara2016, Ceccarelli2018}.
Since dust growth in the ISM is strongly affected by the metallicity in a galaxy \citep[e.g.,][]{Inoue2003, Asano2013a}, they claim that it might not be very important in young galaxies with low metallicity.
However, other studies have shown that the dust mass in distant galaxies cannot be explained without considering metal accretion \citep[e.g.,][]{Pipino2011, Valiante2011, Zhukovska2014, Michaowski2015, Mancini2015, Lesniewska2019, Rouille2020}.

\citet{Asano2013a} defined the critical metallicity, $Z_\mr{cr}$, as the metallicity of the ISM, at which the production rate of dust from stars (AGB and SNe II) become equal to the mass growth rate in the ISM. 
When the metallicity reaches $Z_\mr{cr}$, the dust mass increases suddenly and nonlinearly. 
This rapid increase in dust mass is caused by the following process \citep[e.g.,][]{Hirashita2009, Asano2013a, Asano2013b}.
First, the metal accretion depends on the metallicity and total surface area of the dust grains.
As the dust size increases through the metal accretion, yet another process of dust evolution in the ISM, shattering, is more likely to occur.
This process is basically the collision of grains with each other and redistribute the mass of dust into smaller-sized grains.
When the shattering becomes effective, the total surface area of the dust grain per mass increases, and the metal accretion becomes more efficient.
This cycle leads to the sharp increase of the total dust mass along with metallicity.
Therefore, not only the total dust mass, but also it is of vital importance to take into account the grain size distribution to discuss the evolution of dust in galaxies. 

The absorption and scattering coefficients of dust as a function of wavelength depend on the size and composition of grains.
When a dust grain absorbs light, its temperature rises and the absorbed energy is re-emitted at longer wavelengths (mainly IR). 
The wavelength of the re-emission depends on the instantaneous temperature of the grain, and the temperature strongly depends on the size of the grain.
Thus, as already mentioned, the mass, size, and composition of dust grains play a fundamental role in shaping the SED of a galaxy.
Fitting to the SED of distant galaxies by an empirical dust emission model without evolution may lead to erroneous results.
This happens, for example, when we use a model in which the dust size distribution is constant. 
Recently, some galaxies with a large amount of dust ($M_\mr{dust} > 10^{6}~M_\odot$) have been observed at $z > 6$ \citep[e.g.,][]{Watson2015, Laporte2017, Tamura2019}.
Now it is a proper moment to develop a new SED model based on the theory of dust evolution, after the advent of the Atacama Large Millimeter/Submillimeter Array (ALMA) and other large facilities at this wavelength range.

The dust evolution in the ISM has been considered by a number of previous studies \citep[e.g.,][]{Dwek1998, Calura2008, daCunha2010, Asano2013a, Asano2013b, Asano2014, Mancini2015, Schneider2016, Ginolfi2018, DeVis2017, DeVis2019, Hirashita2019, DeLooze2020, Nanni2020, Burgarella2020}.
We put our basis on the theoretical framework of dust evolution proposed by \citet{Asano2013a, Asano2013b, Asano2014} and \citet{Nozawa2015} (hereafter Asano model) to develop a new radiative transfer SED model.
The Asano model considers SNe II and AGB stars as dust production sources, and not only the metal accretion but also shattering and coagulation as the dust evolution process in the ISM, which enables us to determine the dust mass and size distribution in all galaxy ages from a first principle.
\citet{Hirashita2019} developed a dust evolution model also based on the Asano model, but with a better computational performance to apply to cosmological simulations.
Recall that the Asano model considers different physical quantities (e.g., ambient gas density of SN, hydrogen gas density, and magnetic field strength) for various galaxies to treat the dust destruction by SN and dust collisions.
However,considering dust on the cosmological scale, it is impossible to reach a galaxy scale resolution.
Therefore, \citet{Hirashita2019} adopt many simplifications to optimize their model for the simulations.
In contrast, since we aim at calculating SED of an individual galaxy, we make a maximal use of the Asano model.

There have been several SED models that include the evolution of dust in the ISM.
For example, \citet{Schurer2009} is the SED model which is based on the dust model of \citet{Calura2008}.
They calculate the chemical evolution in a single gas phase and dust evolution including metal accretion.
However, since \citet{Schurer2009} do not consider shattering and coagulation, the rapid increase of the total dust mass does not occur.
Version 3 of \Pegase \citep{Fioc2019} considers the evolution of dust mass, and can calculate not only the radiation from stars but also the extinction by dust grain and the radiation of dust with a stochastic temperature distribution.
\Pegase does not take into account the dust size distribution, and assumes that the fraction and the size distribution of each grain species do not evolve.
In this paper, we construct a new galaxy SED model, including the dust evolution theory proposed by \citet{Asano2013a, Asano2013b, Asano2014}.
We adopt the mega-grain approximation (MGA) with a one-dimensional plain parallel galaxy \citep{Varosi1999, Inoue2005} to make the radiative transfer calculation faster.

This paper is organized as follows.
In Section 2, we introduce how to calculate the SED for each component.
In Section 3, as an example of our SED model, we show the SED of a Milky Way (MW)-like galaxy.
In Section 4, we discuss the effect of parameters on the model SEDs.
Section 5 is devoted to the conclusions.

\section{Methods: Construction of SED model}

To construct a galaxy SED model, we synthesis the stellar SED calculated by version 2 of \Pegase\ \citep[][\Pegase.2]{Fioc1999}, the dust evolution model based on \citet{Asano2013a, Asano2013b, Asano2014}, dust attenuation calculated by radiative transfer with MGA in a one-dimensional galaxy \citep{Varosi1999, Inoue2005}, and the dust emission by a Monte Carlo simulation.
In this section, we present how to calculate each component.
\S\ref{sec:Equation_of_galaxy_evolution} introduces the equation of mass evolution in a galaxy.
\S\ref{sec:Dust_evolution_model} shows the details of the dust chemical evolution model.
\S\ref{sec:Stellar_SED} is the overview of calculation of stellar SED by \Pegase.2.
\S\ref{sec:Dust_properties} and \ref{sec:Radiative_transfer_in_one_dimensional_galaxy} show dust properties with mega-grain approximation and radiative transfer in one-dimensional galaxy, respectively.
In \S\ref{sec:Dust_temperature_distribution} and \ref{sec:Dust_radiation}, we explain the calculation of the dust temperature distribution by Monte Carlo simulation and the dust emission by the distribution.

\subsection{Equations governing galaxy evolution} \label{sec:Equation_of_galaxy_evolution}

We consider stars, gases and dust grains as the components of a model galaxy.
For simplicity, we assume a one-zone galaxy model, where physical quantities vary uniformly over the entire galaxy.
The time evolution of the total stellar mass $M_\ast(t)$, the ISM mass $M_\mr{ISM}(t)$, the metal mass $M_\mr{Z}(t)$, and the dust mass $M_\mr{d}(t)$ at an age of galaxy $t$ are represented as \citep{Lisenfeld1998, Asano2013a},
\begin{align}
\diff{M_\ast(t)}{t} & = \mr{SFR}(t) - R(t), \\
\diff{M_\mr{ISM}(t)}{t} & = -\mr{SFR}(t) + R(t) + \diff{M_\mr{infall}(t)}{t}, \\
\diff{M_\mr{Z}(t)}{t} & = -Z(t)\mr{SFR}(t) + R_\mr{Z}(t) + Y_\mr{Z}(t), \\
\diff{M_\mr{d}(t)}{t} & = -D(t)\mr{SFR}(t) + Y_\mr{d}(t) \notag \\ 
& -\Big(\frac{\mr{d}M_\mr{d}(t)}{\mr{d}t} \Big)_\mr{SN}
+ \Big(\frac{\mr{d}M_\mr{d}(t)}{\mr{d}t} \Big)_\mr{acc}, \label{equ:dMdt_dust}
\end{align}
where $\mr{SFR}(t)$ is the star formation rate, and $R(t)$ and $R_\mr{Z}(t)$ are the mass of the gas and metal taken into stars from ISM and returned to ISM when stars die per unit time, respectively.
$\mr{d}M_\mr{infall}/\dt$ is the infall gas rate, which is assumed to be zero in this paper except \S\ref{sec:Comparison_between_closed-box_and_infall_model}.
$Z(t) \equiv M_\mr{Z}/M_\mr{ISM}$ is the metallicity, and $D(t) \equiv M_\mr{d}/M_\mr{ISM}$ is the mass fraction of dust with respect to the total amount of metal.
$Y_\mr{Z}(t)$ and $Y_\mr{d}(t)$ are the metal and dust masses newly produced by stars per unit time, respectively.
$(\mr{d}M_\mr{d}(t)/\mr{d}t)_\mr{SN}$ and $(\mr{d}M_\mr{d}(t)/\mr{d}t)_\mr{acc}$ is the change of grain mass cased by SN shock and metal accretion.
In the Asano model, three phases are supposed in the ISM: warm neutral medium (WNM, with gas temperature $T_\mr{gas} = 6000$~K, and hydrogen number density $n_\mr{H} = 0.3~\mr{cm^{-3}}$), cold neutral medium (CNM, with $T_\mr{gas} = 100$~K, $n_\mr{H} = 30~\mr{cm^{-3}}$), and molecular cloud (MC, with $T_\mr{gas} = {25}$~K, $n_\mr{H} = 300~\mr{cm^{-3}}$) \citep{Nozawa2015}.
In the MC, dust grains form icy mantle on the grain surface. \citep[e.g.,][]{Kalvans2017, Ceccarelli2018}.
However, since the properties of the icy mantle are not well understood yet, we do not consider its effect in this work.
As for the dust growth in the ISM, metal acceleration occurs only in CNM and MC, and shattering and coagulation occur in all three phases.
Because metal accretion occurs effectively in high density regions, the grain growth in the MC is more prominent.
In this paper, we fix the phase fraction of WNM, CNM, and MC to $\eta_\mr{WNM} = 0.5$, $\eta_\mr{CNM} = 0.3$, and $\eta_\mr{MC} = 0.2$, respectively, the same values to that of \cite{Nozawa2015} used to reproduce the MW extinction curve.
These fractions are constant throughout the calculation of the dust model.
For each time step, the dust grain is redistributed into each ISM phase so that the mass fraction of each ISM phase is maintained.
Thus, we consider the dust grain cycling between the different ISM phases.
Also, we do not consider the outflow effects in this model.
We assume that at the age of $t = 0$, the galaxy contains no stars and dust, and contains only zero-metallicity gas (i.e., $M_\mr{star}(0) = M_\mr{Z}(0) = M_\mr{d}(0) = 0$, and $M_\mr{ISM}(0)$ is total galaxy mass).

We adopt the Schmidt law \citep{Schmidt1959}, $\mr{SFR}(t) \propto M^n_\mr{ISM}$ for SFR with $n = 1$ for simplicity, as
\begin{equation}
	\mr{SFR}(t) = \frac{M_\mr{ISM}(t)}{\tau_\mr{SF}}, \label{equ:SFR}
\end{equation}
where $\tau_\mr{SF}$ is the timescale of star formation.
In this paper, initial galaxy mass $M_\mr{ISM}(0)$ and $\tau_\mr{SF}$ are set to be $10^{11}~\mr{M_\odot}$ and $3$~Gyr as fiducial values.
$R(t)$, $R_\mr{Z}(t)$, and $Y_\mr{d}(t)$ are represented as
\begin{align}
	R(t) & = \int^{100~M_\odot}_{m_\mr{min}(t)}
	[m - \omega(m, Z(t - \tau_m))]
	\phi(m) \notag \\
	& ~~~~~~~~~~~~~~~~~~~~~~~~~~~~~~~~~~~~~~~~~\times \mr{SFR}(t - \tau_m)\,\mr{d}m,\\
	R_\mr{Z}(t) & = \int^{100~M_\odot}_{m_\mr{min}(t)}
	[m - \omega(m, Z(t - \tau_m))]
	\phi(m) \notag \\ 
	& ~~~~~~~~~~~~~~~~~~~~~~~~~~~~~~~~~~~~~~~~~~\times \mr{SFR}(t - \tau_m) Z(t - \tau_m)\,\mr{d}m, \\
	Y_\mr{Z}(t) & = \int^{100~M_\odot}_{m_\mr{min}(t)}
	m_\mr{Z}(m, Z(t - \tau_m)) \phi(m)\mr{SFR}(t - \tau_m)\,\mr{d}m, \\
	Y_\mr{d}(t) & = \int^{100~M_\odot}_{m_\mr{min}(t)}
	m_\mr{d}(m, Z(t - \tau_m)) \phi(m)\mr{SFR}(t - \tau_m)\,\mr{d}m	,
\end{align}
where $m_\mr{min}(t)$ is the lower limit mass of star which can explode at time $t$, $\phi(t)$ is the initial mass function (IMF), $\omega(m,Z(t-\tau_m))$, $m_\mr{Z}(m, Z(t-\tau_m))$, and $m_\mr{d}(m, Z(t-\tau_m))$ are the remnant mass which remains after a star explodes, and the metal mass and the dust mass newly produced by a star of mass $m$ and metallicity $Z(t-\tau_m)$.
As for $\omega$ and $m_\mr{Z}$, we adopt \citet{Ventura2013} for AGB stars with mass $m = 1$--$8~M_\odot$ and metallicity $Z = (0.015$, 0.4)~$Z_\odot$, and \citet{Kobayashi2006} for SNe II with progenitor mass $m = 13$--$40~M_\odot$ and metallicity $Z = (0.0$, 0.05, 0.3, 1.0)~$Z_\odot$.
We interpolate and extrapolate all data tables over mass and metallicity in this paper.
$\tau_m$ is the lifetime of a star with mass $m$ and we use the following equation by \citet{Raiteri1996},
\begin{equation}
	\log \tau_m = a_0(Z) + a_1(Z)\log m + a_2(Z) (\log m)^2,
\end{equation}
with
\begin{align}
	a_0(Z) & = 10.13 + 0.07547 \log Z - 0.008084 (\log Z)^2, \\
	a_1(Z) & = -4.424 - 0.7939 \log Z - 0.1187 (\log Z)^2, \\
	a_2(Z) & = 1.262 + 0.3385\log Z + 0.05417 (\log Z)^2.
\end{align}
This equation was obtained by fitting the calculation result of stars with stellar mass range 0.6--120~$M_\odot$ and metallicity range 0.0004--0.05 by the Padova group \citep{Alongi1993, Bressan1993, Bertelli1994}.
We use the Salpeter IMF \citep{Salpeter1955}:
\begin{equation}
	\phi(m) \propto m^{-2.35}. \label{equ:IMF}
\end{equation}
The IMF is normalized as
\begin{equation}
	\int^{100~M_\odot}_{0.1~M_\odot}
	\phi(m)m~\dm = 1~M_\odot. \label{equ:IMF_norm}
\end{equation}

\subsection{Dust evolution model} \label{sec:Dust_evolution_model}

In this paper, we adopt the Asano model. 
The Asano model takes into account ten species for dust; C, Si, SiO$_2$, Fe, FeS, Al$_2$O$_3$, MgO, MgSiO$_3$, Mg$_2$SiO$_4$, and Fe$_3$O$_4$ \citep{Nozawa2007, Zhukovska2008}.
Since the Asano model calculates the evolution for each dust species, the dust composition evolves with time.
For simplicity in this work, we divide dust grains into two representative families, silicate and carbonaceous grains, for the grain growth and grain-grain collision, because the optical properties of other dust species are not well understood yet.
Further, among the carbonaceous grains, the smaller ones are treated as polycyclic aromatic hydrocarbon (PAH) grains.
The fraction of graphite in carbonaceous dust grain is obtained by the following formula \citep{Draine2007},
\begin{equation}
    f_\mr{gra} =
    \begin{cases}
      0.01 & (a < 50~\mr{\mathring{A}}) \\
      0.01 + 0.99 \left[1 - \left(\frac{50~\mr{\mathring{A}}}{a}\right)^3\right] & (a > 50~\mr{\mathring{A}})
    \end{cases}.
\end{equation}
Where $a$ is the dust grain radius and the fraction of PAHs is defined as $f_\mr{PAH} = 1 - f_\mr{gra}$.
Since the PAHs are divided into ionized and neutral PAHs, which have different optical properties. 
The fraction of ionized PAH is shown in Figure~7 of \citet{Li2001}.
A part of carbonaceous grains could consist of amorphous carbon.
According to \citet{Nozawa2015}, graphite is accepted for reproducing the attenuation curve of nearby galaxy like MW, and amorphous carbon is accepted for galaxies that do not have 2175~\AA\ bumps in the attenuation curve, such as high-$z$ quasars.
Since the differences between these compositions are not well understood yet, we only consider graphite grains in this work. 
It is also possible to take into account amorphous carbon in our SED model.

This model considers only AGB and SNe II as the sources of dust grains.
For simplicity, we do not consider SNe Ia contributions because they are considered to be a very minor contributor to the total mass of dust \citep[e.g.,][]{Calura2008}.

The dust size distribution is represented by the dust number $\fx(a, t)$ and mass $\rhox(\md, t)$ distribution.
$\fx(a, t)\,\da$ and $\rhox(\md, t)\,\dmd$ are the number and mass density of dust grains with radii [$a, a + \da$] and mass [$\md, \md + \dmd$] at time $t$, respectively.
Where 'X' represents the dust species (C: carbonaceous or Si: silicate grain), $a$ is the dust radius, and $\md$ is the dust grain mass.
We assume that the dust grain has a constant density $s$ and is spherical grain, so dust grain mass is
\begin{equation}
\md = \frac{4}{3}\pi a^3s. \label{equ:grain_mass}
\end{equation}
The relation of dust number and mass density is expressed as 
\begin{equation}
    \rhox(\md, t)\,\dmd = \md \fx(a, t)\,\da.
\end{equation}

In the initial condition of the galaxy, the dust mass of all sizes is set to be zero.
Therefore, in the first time step of the computation, only the dust grain is produced by the stars.
In the next time step, the dust size distribution produced by the stars evolves through the dust evolution process in the ISM.
In the following, we explain the details of the Asano model.

\subsubsection{Dust production by AGB} \label{sec:Dust_production_by_AGB}

AGB stars are the final phase of the evolution of low and intermediate mass stars ($< 8~M_\odot$).
They have the carbon-oxygen core and are burning hydrogen and helium surrounding the core.
They release heavier elements in the ISM, and dust grains are formed in the ejecta. 
The dust size distribution produced by AGB stars depends on the progenitor mass and suggested that it is represented by a log-normal distribution with a peak at $\sim 0.1~\mathrm{\mu m}$ by \citet{Winters1997}.
Further, \citet{Yasuda2012} calculates dust formation by AGB stars with the hydrodynamical simulation including SiC production.
They suggest that the mass distribution per unit logarithmic bin $a^4f(a)$ is described by a log-normal distribution with a peak at 0.2--0.3~$\mu$m.
We assume the size distribution of dust grains from AGB stars is represented by the  log-normal with a peak at 0.1~$\mu$m with a standard deviation of $\sigma = 0.47$.
This shape reproduces Figure 7 of \citet{Yasuda2012}.
We assume the same size distributions for all dust species.

As for the dust mass produced by AGB stars, we adopt \citet{Ventura2012, Ventura2013}.
They consider AGB stars with a mass range of 1--8~$M_\odot$ and metallicity range of $Z = (0.05$, 0.4)~$Z_\odot$.
The condensation fraction of the key elements is $\sim 0.3$ for silicate with progenitor stellar mass $M_\mr{AGB} = 6~M_\odot$ and initial metallicity $Z = 0.05~Z_\odot$, and $\sim 0.05$ for carbon with $M_\mr{AGB} = 3~M_\odot$ \citep{Ventura2012}.
We interpolate and extrapolate their data to obtain the dust yield at the required stellar mass and metallicity, as in \S\ref{sec:Equation_of_galaxy_evolution}.

\subsubsection{Dust production by SNe II} \label{sec:Dust_production_by_SNeII}
Massive stars end their lives as supernovae (SNe), and dust grains are formed in the ejecta of the SNe.
The element synthesis determines the dust grains composition in stars and the mechanism of explosion \citep{Nozawa2003}.
Furthermore, the reverse shock of SN destroys the dust grains by sputtering \citep{Nozawa2007, Bianchi2007}. 
Dust destruction by SN reverse shock is still under debate and there is no common agreement \citep[e.g.,][]{Gall2014, Biscaro2016, Matsuura2019, Slavin2020}, but we use \cite{Nozawa2007} as a working hypothesis in this paper.
\citet{Nozawa2007} calculates the grain size distribution produced by SNe II, and we adopt it for SNe in a progenitor mass range of 13--30~$M_\odot$.
They calculate only production by SNe II from zero-metallicity star, but the size distribution and composition of dust produced by the SNe II are less dependent on the metallicity of the progenitor stars \citep[e.g.,][]{Todini2001, Kozasa2009}.
Therefore, we assume that the dust production from the SN does not depend on metallicity and uses the stellar mass interpolated and extrapolated.

\citet{Nozawa2007} discusses two extreme cases for the structure of the cores of progenitor stars, mixed and unmixed.
According to \citet{Hirashita2005b}, the unmixed model gives a better fit to the observed high-$z$ extinction curve of SDSS J1048+4637 at $z = 6.2$ \citep[][]{Maiolino2004}.
Thus, we adopt the  unmixed model in this paper.
The condensation fraction is about 0.003--0.006 \citep{Nozawa2007}.

\subsubsection{Dust destruction by supernova shock}
\label{sec:Dust_destruction_by_supernova_shock}

Dust grains in the ISM are partially destroyed by SN shocks \citep[e.g.,][]{Jones1996, Nozawa2006}.
The SN shocks decrease the total dust mass, and change the size distribution through the sputtering process \citep{Nozawa2006}.
The sputtering is separated into thermal and non-thermal.
The thermal sputtering is caused by the motion of hot gas, and nonthermal one is caused by relative motion between gas and dust grain.
The sputterings depend on grain size, gas density, temperature (for thermal sputtering), and the relative velocity between dust grain and gas (for nonthermal sputtering).

We adopt the result by \cite{Yamasawa2011} for the treatment of the SN destruction.
The grain number density after the destruction by SN shocks, $\fx^\prime(a, t)$, is formulated as
\begin{equation}
  \fx^\prime(a, t) = \int^{a_\mr{max}}_a
  \eta_\mr{X}(a, a^\prime)
  \fx(a^\prime, t)\,\da^\prime. \label{equ:fx_sn}
\end{equation}
Where $\eta_\mr{X}(a, a^\prime)$ is the conversion efficiency of SN sputtering defined as the conversion rate of dust grains from radii $[a, a + \da]$ to $[a^\prime, a^\prime + \da^\prime]$.
$a_\mr{max}$ is the maximum radius of dust grain and we adopt $a_\mr{max} = 8~\mr{\mu m}$ \citep{Asano2013b}.
This value is large enough to represent the maximum size produced by shattering and coagulation \citep{Hirashita2009}.
\cite{Yamasawa2011} calculate $\eta_\mr{X}$ by the method developed by \cite{Nozawa2006}.
In this process, the size of dust grains is only reduced by destruction, if $a > a^\prime$, $\eta_\mr{X} = 0$.
Equation~(\ref{equ:fx_sn}) represents the increasing amount of dust with radii $[a, a + \da]$ by SN destruction of dust larger than $a$.
The actual upper limit of the integration corresponds to the maximum dust size of the distribution before the shock passes.
The change of grain number density caused by SN shock is represented as
\begin{align}
\mr{d}\fx(a, t) & = \fx^\prime(a, t) - [1 - \eta_\mr{X}(a,a)]\fx(a, t) \notag \\
 & = \int_0^{a_\mr{max}} \eta_\mr{X}(a,a^\prime)\fx(a^\prime, t)\,\da^\prime - \fx(a, t). \label{equ:df_SN}
\end{align}

The change of grain mass density by SN shock at a grain radius $a$ and the time $t$ is represented by Eq.~(\ref{equ:df_SN}) as,
\begin{align}
  \left(\diff{\rhox(\md, t)}{t}\right)_{SN}
  = & \md\diff{\fx(a,t)}{t} \notag \\
  = & -\tau_\mr{SN,X}^{-1}
     \Big[ \rhox(\md, t) \notag \\
     & - \md\int^{a_\mr{max}}_0 \eta_\mr{X} (a, a^\prime) \fx(a^\prime, t)\,\da^\prime \Big].
\end{align}
If we integrate this equation with respect to $a$ and summing up the dust species, it agrees with the third term of the right hand side of Equation (\ref{equ:dMdt_dust}).
The timescale of dust destruction $\tau_\mr{SN}(t)$ by SN is expressed as,
\begin{equation}
	\tau_\mr{SN}(t) = \frac{M_\mr{ISM}(t)}{\epsilon m_\mr{swept}\gamma_\mr{SN}(t)}, \label{equ:tau_SN}
\end{equation}
where $\epsilon$ is the efficiency of the dust destruction by SN shocks, and $\gamma_\mr{SN}(t)$ is the SN rate.
The SN rate is expressed as follows by \citep{McKee1989b, Nozawa2006},
\begin{equation}
	\gamma_\mr{SN}(t) = \int^{40~M_\odot}_{\max(m_\mr{min}(t), 8~M_\odot)}
	\phi(m) \mr{SFR}(t - \tau_\mr{m}) \mr{d}m. \label{equ:SN_rate}
\end{equation}
The integration range is determined by when the SNe can occur \citep{Heger2003}.
When $t < \tau(40~M_\odot)$, $\gamma_\mr{SN}(t) = 0$.
We assume $\epsilon = 0.1$ \citep{McKee1989b, Nozawa2006}.

$m_\mr{swept}$ is the ISM mass swept by a SN shock.
$m_\mr{swept}$ depends on the density and metallicity of the ISM \citep{Nozawa2006, Yamasawa2011}.
When the ISM density is high, $m_\mr{swept}$ is small because there are many particles that slow down the SN shock.
When the metallicity is high, efficient line cooling with metal results in a faster shock deceleration and smaller $m_\mr{swept}$.
We use the following formulae fitted by \citet{Yamasawa2011},
\begin{equation}
	m_\mr{swept} = 1535n^{-0.202}_\mr{SN}
	\left[ \left(
	Z/Z_\odot
	\right)
	+ 0.039
	\right]^{-0.289} ~M_\odot, \label{equ:swept_mass}
\end{equation}
where $n_\mr{SN}$ is the ISM density surrounding SNe.
The fitting accuracy is within 16\% for $0.03~\mr{cm}^{-3} \leq n_\mr{SN} \leq 30~\mr{cm}^{-3}$ and for $10^{-4} \leq Z/Z_\odot \leq 1.0$ \citep{Yamasawa2011}, and we use $n_\mr{SN} = 1.0~\mr{cm^{-3}}$.

\subsubsection{Grain growth by metal accretion} \label{sec:Grain_growth}

In the cold phase of the ISM, the metal in the gas phase is accreted onto the pre-existing dust grain surface, which increases the radius of the dust grain and the total mass of the dust, known as grain growth \citep[e.g.,][]{Dwek1980, Draine2009a, Jones2011}.
We assume that grain growth occurs in the CNM ($T_\mr{gas} = 100$~K and $n_\mr{H} = 30~\mr{cm^{-3}}$) and the MC ($T_\mr{gas} = 25$~K and $n_\mr{H} = 300~\mr{cm^{-3}}$). The total mass fraction in these ISM phases is assumed to be 0.5 \citep{Nozawa2015}.
We treat only refractory grains (silicate and carbonaceous dust) and do not consider the icy mantle.
We assume that a grain instantly becomes a sphere with a smooth surface, and we adopt only the geometric cross-section, i.e., we do not consider the effect of the Coulomb interaction.
Here, we consider only two species of dust grain, carbonaceous with key element C and silicate with Si.
\citet{Jones2011} indicate that in an H$_2$, CO and H$_2$O-rich environment, Si, Fe and Mg accretion forms silicates through complex chemical reactions such as ice formation.
However, they also say that the spectrum of silicates formed in such a scenario is inconsistent with actual observations.
In this paper, we do not know the chemical properties well, so it is assumed that only the same key element X accretes onto the dust of key elements X.
Since grain growth requires a sufficient amount of metals and dust grains in the ISM, grain growth is difficult to occur in a very young galaxy, but it becomes efficiently processed at 1~Gyr in general \citep{Asano2013a}.

In the following, we introduce the formalism of size evolution by \cite{Hirashita2011}.
The collision rate at which an atom of element X with radius $a$ with the surface of the dust grain is expressed as follows \citep{Evans1994}: 
\begin{equation}
  \mathcal{R} = \pi a^2 n_\mr{X}(t) v_\mr{th}, \label{equ:collision_rate}
\end{equation}
where $n_\mr{X}(t)$ is the number density of the key element X in gas phase and $v_\mr{th} $ is the thermal velocity
\begin{equation}
      v_\mr{th} = \left(\frac{8kT_\mr{gas}}{\pi m_\mr{X}}\right)^{1/2},
\end{equation}
where $k$ is the Boltzmann constant, $T_\mr{gas}$ is the gas temperature and $m_\mr{X}$ is the atomic mass of the key element X.
In reality, since metals other than the corresponding key element may accrete onto the grain surface, Equation (\ref{equ:collision_rate}) represents the accretion rate associated with the key element.
The evolution of grain mass $\mr{d}\md(a, t)/\mr{d}t$ is
\begin{equation}
  \diff{\md(a, t)}{t} = g_\mr{X}^{-1}\md\alpha_\mr{acc}\mathcal{R}, \label{equ:dm_dt}
\end{equation}
where $g_\mr{X}$ is the mass fraction of the key element X in a specific grain species (silicate: 0.166, graphite: 1.00).
We assume Mg$_{1.1}$Fe$_{0.9}$SiO$_{4}$ for the composition of silicate \citep{Draine1984}.
$\alpha_\mr{acc}$ is the sticking probability of atoms that collide with grains.
It is very difficult to quantify whether the sticking atoms become a part of the grain \citep[e.g.,][]{Jones2011}.
This value may be almost 1 in the low temperature environment \citep[e.g.,][]{Zhukovska2008}, so this paper sets $\alpha_\mr{acc} = 1$ for simplicity.

$n_\mr{X}(t)$ is estimated as
\begin{equation}
  n_\mr{X}(t) = 
  \frac{\rho_\mr{ISM}^\mr{eff}}{\md}
  \frac{M_\mr{X}(t) - g_\mr{X}M_\mr{d,X}(t)}{M_\mr{ISM}(t)}, \label{equ:n_grain_growth}
\end{equation}
where $M_\mr{X}(t)$ is the total mass of element X (including gas and dust), $M_\mr{ISM}(t)$ is the total mass of gas, $M_\mr{d,X}(t)$ is the dust mass associated with element X, and $\rho^\mr{eff}_\mr{ISM}$ is the effective ISM mass density which is averaged mass density of the cloud where accretion process occurs.
$\rho^\mr{eff}_\mr{ISM}$ is calculated as $\rho_\mr{ISM}^\mr{eff} = \mu m_\mr{H}n_\mr{H,acc}$, where $\mu = 1.4$ is the mean atomic weight, $m_\mr{H}$ is the hydrogen atom mass, and $n_\mr{H,acc}$ is the mean hydrogen number density in the ISM where the accretion process takes place.
When $\eta = \eta_\mr{CNM} + \eta_\mr{MC} =0.5$, $n_\mr{H,acc}$ is 130 $\mr{cm^{-3}}$.
The second term on the right hand side represents the gas mass of element X to total ISM mass ratio.
From (\ref{equ:collision_rate})--(\ref{equ:n_grain_growth}), the grain mass growth rate is
\begin{align}
\diff{\md(a,t)}{t}
  = & \frac{\pi a^2 \md \alpha_\mr{acc} v_\mr{th}}{g_\mr{X}} n_\mr{X}(t) \notag \\
  = & \frac{\pi a^2 \alpha_\mr{acc}\rho_\mr{ISM}^\mr{eff}     v_\mr{th}}{g_\mr{X}}
  \frac{M_\mr{X}(t) - g_\mr{X}M_\mr{d,X}(t)}{M_\mr{ISM}(t)}. \label{equ:grain_mass_growth_rate}
\end{align}
The total mass growth rate by accretion process $\left( \mr{d}M_\mr{d}(t)/\mr{d}t\right)_\mr{acc}$ is calculated by integrating the Equation (\ref{equ:grain_mass_growth_rate}) with respect to the grain radius and summing up for all dust species in all ISM phases.
From Equation (\ref{equ:grain_mass}) and (\ref{equ:grain_mass_growth_rate}), grain radius growth rate is represented as
\begin{equation}
\diff{a}{t}
  = \frac{\alpha_\mr{acc}\rho_\mr{ISM}^\mr{eff}     v_\mr{th}}{4sg_\mr{X}}
  \frac{M_\mr{X}(t) - g_\mr{X}M_\mr{d,X}(t)}{M_\mr{ISM}(t)}. 
\end{equation}
In computation, we solve this equation for each size bin at each time step.
By transferring all the dust that was in the radius bin before metal accretion to the radius bin after accretion, the evolution of size distribution by metal accretion can be calculated.
The number of dust grains does not change in this process.

\subsubsection{Grain-grain collision} \label{sec:Grain-grain_collision}

We consider two types of gain-grain collisions, shattering, and coagulation.
They only change dust size distribution and conserve total grain mass in the ISM.
Which processes occur is determined by the relative velocity between two collisional grains.
In the case that the relative velocity is fast, shattering is easy to occur.
On the contrary, when the relative velocity is small, coagulation occurs.
Relative velocities between dust grains can be caused by ubiquitous ISM turbulence \citep[e.g.,][]{Draine1985, Ossenkopf1993, Lazarian2002, McKee2007, Hirashita2009, Ormel2009}.
Furthermore, because the dust is thought to be magnetized \citep{Arons1975}, it is necessary to consider the motion of grains due to magnetohydrodynamics (MHD) turbulence.
We consider dust collisions by applying the relative velocity of grains in MHD turbulence calculated by \citet{Yan2004}.
The velocity is calculated in consideration of gas drag (hydro drag) and gyroresonance.
When a grain with a relative velocity higher than the threshold collides, it is shattered into small pieces.
Since the larger grains are affected by turbulence strongly, they have a large relative velocity and it can occur shattering.
\cite{Yan2004} indicate that grains with $a > 10^{-6}$~cm can accelerates to velocity (1--2~$\mr{km\,s^{-1}}$) close to the shattering threshold in CNM.
In WNM, gyroresonance accelerates grain with $a > 2$--$3 \times 10^{-5}~\mr{cm}$ to high relative velocity ($\sim 20~\mr{kms^{-1}}$).
On the contrary, small dust has a small relative velocity and causes coagulation.
When the relative velocity is small, the collision cross section is small, so a high density environment (e.g., MC) is required for coagulation.
We consider silicate and graphite as grain species while grain-grain collision, and they collide only with the same species.
Furthermore, We treat spherical grains which have a constant density.
\citet{Hirashita2009} calculates in various ISM phases (including CNM, WNM, and MC) by shattering and coagulation, and the Asano model applied the same method.

We consider four types of grain-grain collisions, in other words, the relative velocity is divided into four types.
This treatment is the same as \citet{Jones1994} and \citet{Hirashita2009}.
Considering the shattered and coagulated grains are radii $a_1$ and $a_2$, which are called grain 1 and 2, respectively.
The mass of grain 1 and 2 is denoted as $m_1$ and $m_2$.
The relative collisional velocities between grain 1, and 2 are as follows,
 \begin{itemize}
 	\item front collision ($v_{1,2} = v_1 + v_2$)
 	\item back-end collision ($v_{1,2} = |v_1 - v_2|$)
 	\item side collision ($v_{1,2} = v_1$)
 	\item another side collision ($v_{1,2} = v_2$)
 \end{itemize}
where $v_1$ and $v_2$ are the velocity of grain with radii $a_1$ and $a_2$, respectively.
We assume that collisions in all directions have the same probability.

\citet{Jones1996} suggests that the shattering is significantly affecting the grain size distribution in the ISM.
The time evolution of grain mass density by shattering process is 
\begin{align}
	\left[ \diff{\rhox(\md, t)}{t} \right]_\mr{shat}
  & =  -\md\rhox(\md,t) \notag  \\ 
  & \times \intamm \alpha \left[\md, m_1\right]\rhox(m_1,t)~\dm_1 \notag \\
  & + \intamm \intamm \alpha \left[m_1, m_2\right] m^{1,2}_\mr{shat}(\md) \notag \\
  & \times \rhox(m_1,t)\rhox(m_2,t)~\dm_1\dm_2, \label{equ:shattering}
\end{align}
where $\alpha[m_1,m_2]$ is the collision frequency normalized by two grain masses and grain number density and expressed as
\begin{equation}
  \alpha[m_1, m_2] =
  \begin{cases}
    0 & (v_{1,2} < v_\mr{shat}) \\
    \frac{\sigma_{1,2}v_{1,2}}{m_1 m_2}
    & (v_{1,2} > v_\mr{shat})
  \end{cases},
\end{equation}
$m_\mr{shat}^{1,2}(\md)$ represents the total mass of fragments with masses between $\md$ and $\md + \dmd$ as the result of collision between grain 1 and 2.
We assume that the distribution of shattered fragments is proportional to $a^{-3.3}$ \citep{Hellyer1970, Jones1996}.
$\sigma$ is the collisional cross-section and represented as 
\begin{equation}
  \sigma_{1,2} = \beta \pi(a_1 + a_2)^2, \label{equ:cross-section}
\end{equation}
$\beta$ is the coefficient connecting the cross-section and the geometric cross-section, assumed $\beta = 1$ for simplicity.
$v_\mathrm{shat}$ is the threshold of shattering, we assume $1.2~\mathrm{km\,s^{-1}}$ and $2.7~\mr{km\,s^{-1}}$ for silicate and graphite grains, respectively \citep{Jones1996}.
$a_\mr{min}$ and $a_\mr{max}$ are minimum and maximum radius and we adopt $a_\mr{min} = 0.0003~\mr{\mu m}$ and $a_\mr{max} = 8~\mathrm{\mu m}$, respectively \citep{Asano2013b}.
The minimum grain radius in the ISM is less well understood,  even if $a_\mr{min} = 0.001~\mr{\mu m}$, the dust size distribution does not change significantly \citep{Hirashita2012}.

The first term on the right hand side of Equation (\ref{equ:shattering}) represents the decrease of grain mass $\md$ due to destruction by the collisions with other grains.
The second term represents the grain mass $\md$ increase due to the fragments resulting from the collision between grain 1 and 2.
Shattering does not produce larger fragments than the original grain, so it only contributes if either grain 1 or 2 is heavier than $\md$.

The coagulation occurs when the relative velocity is low.
The time evolution for coagulation is expressed as a similar form to shattering,
\begin{align}
  \left[ \diff{\rhox(\md,t)}{t} \right]_\mr{coag}
  & = -\md\rhox(\md, t) \notag \\
  & \times \intamm \alpha \left[\md, m_1\right] \rhox(m_1, t)~\dm_1 \notag \\
   & + \intamm \intamm
    \alpha[m_1, m_2] m_\mr{coag}^{1,2}(\md) \notag \\
   & \times \rhox(m_1, t) \rhox(m_2, t)~\dm_1\dm_2. \label{equ:coagulation_rho}
\end{align}
and
\begin{equation}
	\alpha[m_1,m_2] =
	\begin{cases}
		\frac{\sigma_{1,2}v_{1,2}}{m_1 m_2} & (v_{1,2} < v_\mr{coag}) \\
		0 & (v_{1,2} > v_\mr{coag})
	\end{cases},
\end{equation}
where $m_\mr{coag}^{1,2}$ is the total mass of coagulated grains:
\begin{equation}
	m_\mr{coag}^{1,2}(\md) =
	\begin{cases}
		\md & \mr{when}~\md \leq m_1 + m_2 < \md + \dmd)\\
		0 & \mr{otherwise}
	\end{cases}.
\end{equation}
We use Equation (\ref{equ:cross-section}) as the collisional cross-section for coagulation.
$v_\mr{coag}$ is the threshold velocity of coagulation, and grains with higher relative velocity do not stick.
\citet{Chokshi1993} calculate the threshold velocity as $10^{-3}$--$10^{-1}~\mr{km\,s^{-1}}$ and it depends on the grain size.
Here we assume that the dust grain is a smooth sphere, but in the real picture the grain is fluffy \citep{Ossenkopf1993}.
It has been suggested that the coagulation threshold relative velocity is higher because fluffiness increases the cross-section of grain collisions \citep{Ormel2009, Hirashita2013}.
In addition, \citet{Asano2014} indicate that the coagulation threshold suppresses the production of large grain, producing smaller grain ($a < 0.01~\mr{\mu m}$) than the \citet{Mathis1977}.
Therefore, in this paper, it is assumed that coagulation can occur at all relative velocities without setting the coagulation threshold.
The first term of Equation (\ref{equ:coagulation_rho}) indicates the decrease of the grain with mass $\md$ by coagulation with other grains.
The second term indicates the increase of the grain with mass $\md$ by coagulation between grain 1 and 2.
Since coagulation works effectively on small size dust, it becomes effective after shattering becomes effective and small dust increases.
Coagulation shifts the dust size distribution to the larger one.

\subsubsection{Result of dust evolution model}
\label{sec:Result_of_dust_evolution_model}

\begin{figure}
    \includegraphics[width=\columnwidth]{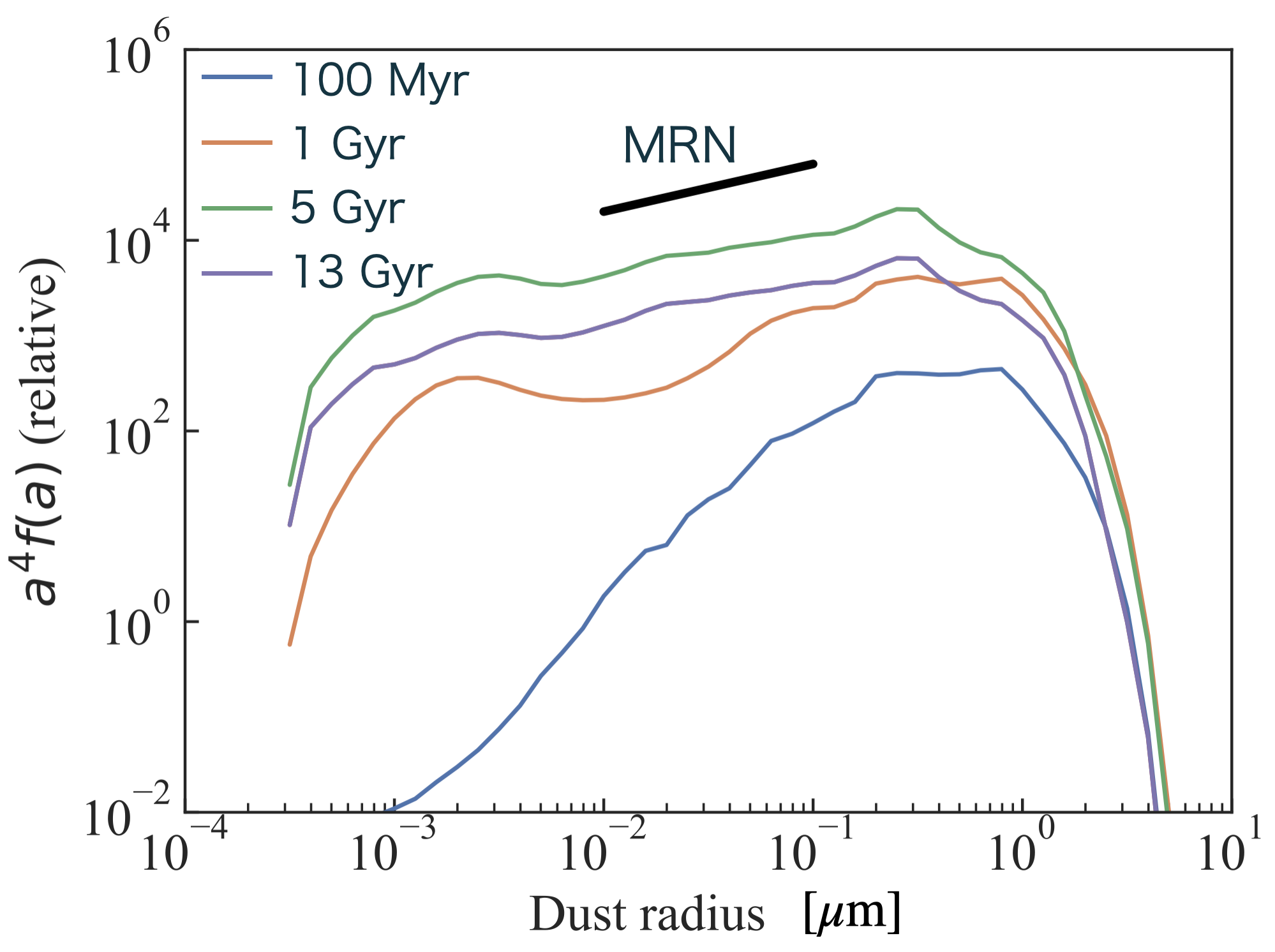}
    \caption{The time evolution of all species of dust grain size distribution.
    Blue, orange, green, and purple curves indicate the age of 100~Myr, 1~Gyr, 5~Gyr, and 13~Gyr, respectively.
    Black line represents the slope of the MRN distribution.
    }
	\label{fig:dust_size_distribution}
\end{figure}

We show the dust grain (including carbon and silicate) size distribution calculated with the star formation timescale $\tau_{SF} = 3$~Gyr and total galaxy mass $10^{11}~M_\odot$ in Figure \ref{fig:dust_size_distribution}.
We note that the total galaxy mass is merely a normalization for our dust model, and can be rescaled freely.

In this representation, the ratio of ISM phases are $\eta_\mr{WNM} = 0.5$, $\eta_\mr{CNM} = 0.3$, $\eta_\mr{MC} = 0.2$.
Blue, orange, green, and purple curves indicate the age of 100~Myr, 1~Gyr, 5~Gyr, and 13~Gyr, respectively.
Black curve indicates the slope of the grain size distribution suggested by \cite{Mathis1977} which reproduces the MW extinction curve.
This is known as the MRN distribution, expressed by a single power law,
\begin{equation}
    f(a)\da \propto a^{-3.5} \da ~ (0.005~\mr{\mu m} < a < 0.25~\mr{\mu m}).
\end{equation}

The overview of the time evolution of dust size distribution is as follows.
\begin{itemize}
    \item $<100$~Myr
    \begin{itemize}
        \item Dust production from SNe dominates, and the original size distribution of SN dust is reflected in the overall dust size distribution.
    \end{itemize}
    \item 100~Myr--1~Gyr
    \begin{itemize}
        \item Metal accretion dominates the evolution of dust.
        \item Shattering and coagulation become effective, and consequently dust mass rapidly increases because of the increase of the total surface area per dust mass.
        \item Production of PAHs is dominated by shattering.
    \end{itemize}
    \item 1--5~Gyr
    \begin{itemize}
    \item Shattering and coagulation become more effective.
    \item Metal accretion also becomes more effective thanks to the increased amount of small dust continuously generated by shattering.
    \item The increase of dust mass is the most rapid between 1~Gyr and 2~Gyr.
    \end{itemize}
    \item 5--13~Gyr
    \begin{itemize}
    \item The production of dust grain by stars decreases, and shattering and coagulation dominate the evolution of the grain size distribution.
    \end{itemize}
\end{itemize}

\begin{figure}
    \includegraphics[width=\columnwidth]{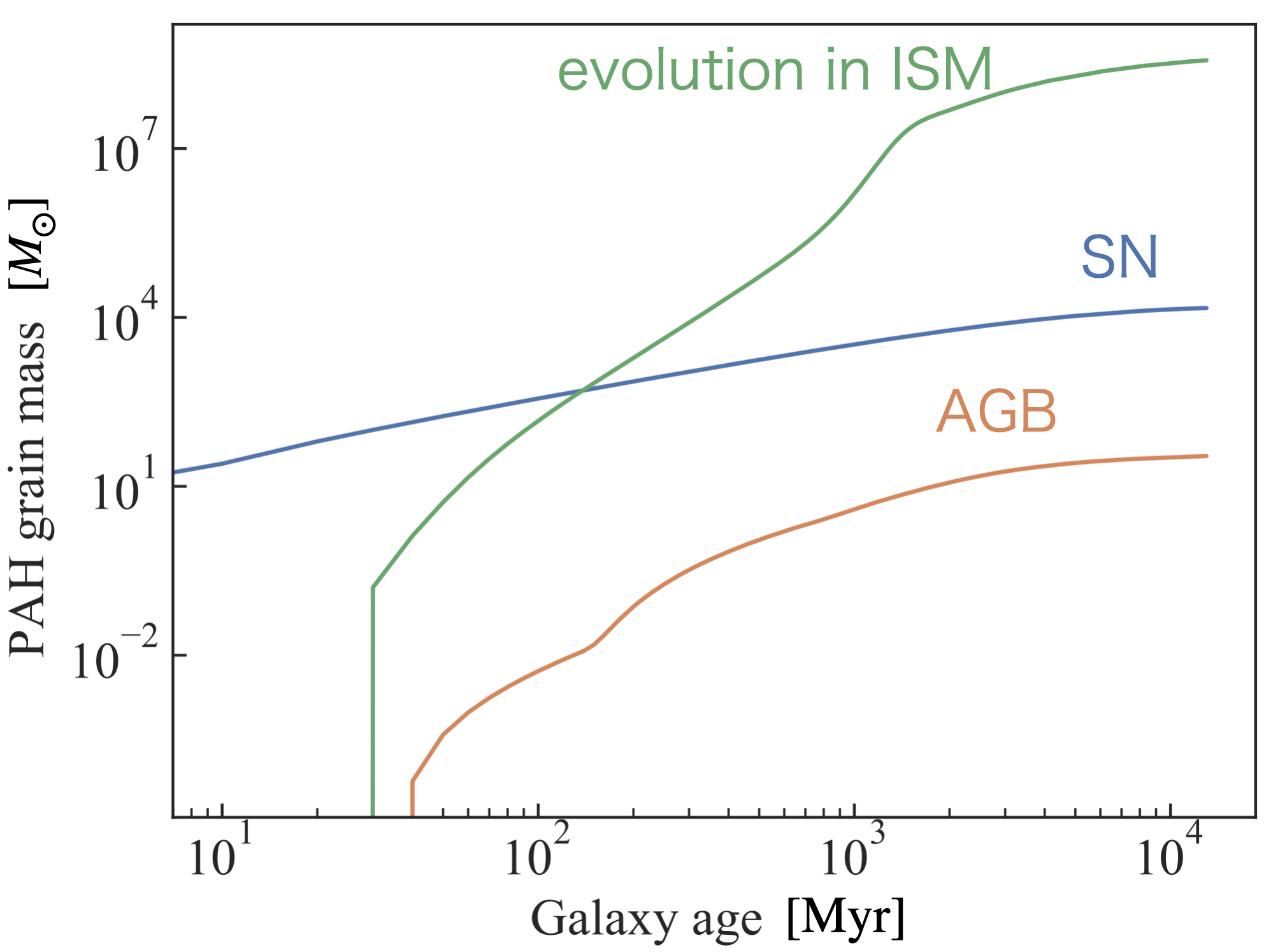}
    \caption{The evolution of PAH mass per each production source.
    The total galaxy mass is $10^{11}~M_\odot$.
    Blue, orange, and green curves represent PAH mass produced by SN, AGB, and evolution in the ISM, respectively.}
	\label{fig:PAH_contribution}
\end{figure}

Details of each step in the evolution of dust size distribution is explained as follows.
At the age of 100~Myr, only a tiny amount of grains exist in the ISM, and the slope of size distribution is completely different from the slope of the MRN distribution.
Particularly, PAHs are not produced in early galaxies.
Figure~\ref{fig:PAH_contribution} shows how different processes (i.e., production by AGB and SNe, and grain growth in the ISM) contribute to the increase of the total PAH mass.
We note that the contribution of destruction processes including the SN shock and astration are not shown here.
The dust destruction process only depends on the grain size and species, then it works in the same way for dust from any production source. 
Therefore, even if dust reduction is taken into account, the ratio of dust mass for each source remains the same.
In galaxies younger than 100~Myr, short-lived SNe II (lifetime $\sim 10^6$--$10^7$~yr) is the main source of dust supply \citep[e.g.,][]{Maiolino2004, Hiraki2008}, as the age of the galaxy is too young for stars to evolve into AGB (lifetime $\sim 10^8$--$10^9$~yr) \citep[e.g.,][]{Morgan2003, Marchenko2006}.
However, smallest grains such as PAHs are supplied from SN stars, though the amount is very small \citep[e.g.,][]{Nozawa2007}.
As the chemical evolution proceeds in the galaxy, the amount of metal in the ISM increases.
Very young galaxies ($\mbox{age} \simeq 20$~Myr) have only a small supply of dust from SNe.
When the galaxy age reaches $\sim 100$~Myr, the smallest grains (PAH) are gradually formed by shattering, and the mass of the PAH increases.
The galaxy must evolve to reach the critical metallicity for dust growth to work effectively \citep{Inoue2011, Asano2013a}.
Since AGB provides larger size dust grains ($>0.1~\mr{\mu m}$), their contribution to PAHs is not significant \citep{Winters1997, Yasuda2012}.

\begin{figure}
    \includegraphics[width=\columnwidth]{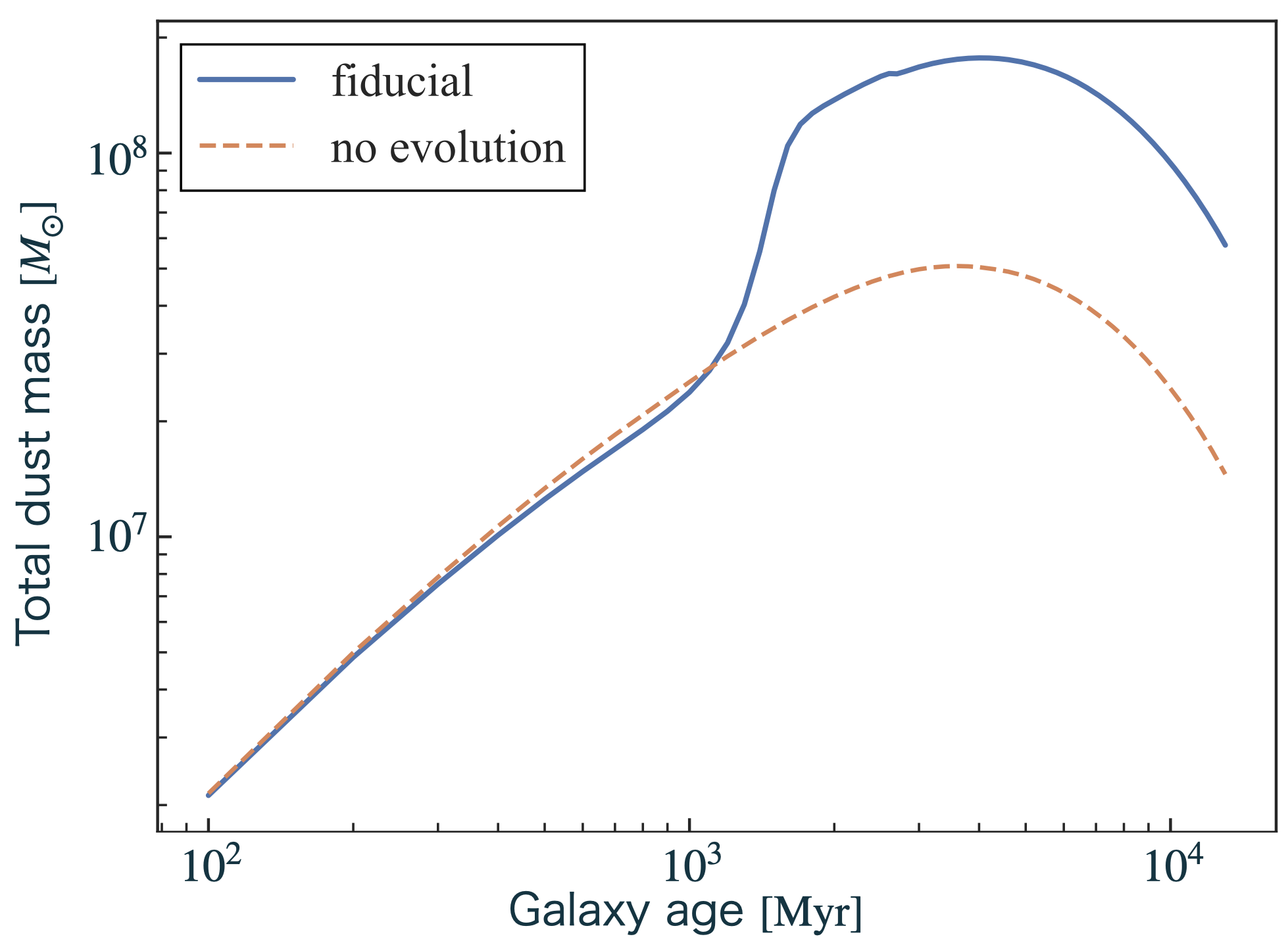}
    \caption{The evolution of dust to gas mass ratio calculated by the Asano model.}
	\label{fig:dust_mass_evolution}
\end{figure}

At 1~Gyr, the total dust mass continues to increase gradually, while the PAH mass starts to increase significantly, because the shattering in the ISM becomes effective.  
The bump in 10$^{-3}$--10$^{-2}~\mr{\mu m}$ in Figure~\ref{fig:dust_size_distribution} is the consequence of the activated shattering process.
We show the evolution of the total dust mass of the model galaxy in Figure~\ref{fig:dust_mass_evolution}.
Solid and dashed lines represent the dust grain evolution in the ISM (fiducial) and without evolution (no evolution) case, respectively.
Figure~\ref{fig:dust_mass_evolution} clearly demonstrates that, if dust grains evolve in the ISM, dust mass rapidly increases by metal accretion in 1--2~Gyr.
It has been suggested that in the MW-like galaxy model, when the metallicity exceeds 0.1~$Z_\odot$, the metal accretion process becomes effective and the dust mass drastically increases \citep{Asano2013a}.
When the metal accretion becomes effective, dust collisions with each other in the ISM become more likely to occur, and the shattering and coagulation also become effective.
The shattering process results in a significant increase in the amount of small dust grains including PAH.
Since the metal accretion depends on the total surface area of dust grain (Equation~(\ref{equ:collision_rate})), the shattering promotes the accretion.
Such a dust growth cycle causes the dust mass to increase nonlinearly.
This cycle is effective between 1~Gyr and 2~Gyr in this model galaxy.

After that, the mass of dust increases and peaks at $\sim 3$~Gyr.
This peak time depends on the timescale of star formation $\tau_\mr{SF} = 3$~Gyr.
After 3~Gyr, the dust mass decreases due to the destruction by SN shocks.
The smaller the dust size, the more effectively the SN destruction works \citep{Nozawa2006}.
In addition, production of dust from stars also decreases due to the decrease of the SFR.
Thus, in total, the dust mass gradually decreases by SN shock and astration.
As the production of dust by stars decreases, coagulation dominates the evolution of dust size distribution.
Due to the SN shock destruction and coagulation, the dust size distribution is biased toward larger radius.
For galaxies with fully grown dust after 5~Gyr, the dust grain size distribution finally converges to a similar function to that obtained from observations of nearby galaxies such as \citet{Schurer2009}. 
A galaxy with the age in $5\mbox{--}13$~Gyr has a dust distribution with a power-law slope similar to the MRN.

In contrast, for the no evolution case, the total grain mass does not increase rapidly and only increases by stellar production with a constant rate up to $\tau_\mr{SF} = 3$~Gyr.
In the age of the $< 1$~Gyr galaxy, no evolution case has a larger grain mass than the mass of the fiducial case.
This is because the no evolution case does not consider the destruction of dust due to SN shocks.
After 3~Gyr, the dust mass decreases by astration.
As described above, if the evolution of dust in the ISM is not taken into account in the calculation, a rapid increase of dust mass in 1--2~Gyr does not appear.

\subsection{Stellar SED} \label{sec:Stellar_SED}

We use the version 2 of \Pegase\ \citep[][\Pegase.2]{Fioc1999} to produce stellar SEDs.
\Pegase\ calculates the stellar emission by stellar population synthesis (SPS) method with simple stellar populations (SSPs).
The SSP represents the time variation of the SED of a single contemporaneous stellar population with a single metallicity and abundance pattern.
The monochromatic luminosity per unit wavelength of SSP is expressed as 
\begin{equation}
	L_\lambda^{\mr{SSP}}(t, Z)
	= \int^{m_\mr{max}}_{m_\mr{min}}
	L_\lambda^{\mr{star}}
	(T_\mr{eff}(t, m), \log g(t, m), Z)\phi(m)~\dm,
\end{equation}
where $L^\mr{star}_\lambda$ is the monochromatic luminosity of a star with the mass in the interval effective temperature $T_\mr{eff}$, surface gravity of stellar $g$, metallicity $Z$, and an age of galaxy $t$ \citep[e.g.,][]{Conroy2013}.
$m_\mr{max}$ and $m_\mr{min}$ are the upper and lower limit of stellar mass, set to be 100~$M_\odot$ and 0.1~$M_\odot$, which is the same as the IMF integration range.
The effective temperature $T_\mr{eff}(t,m)$ and the surface gravity $\log g(t,m)$ are from the stellar evolutionary track.
\Pegase.2 uses the evolutionary track based on the Padova tracks \citep{Bressan1993, Fagotto1994a, Fagotto1994b, Fagotto1994c, Girardi1996}.
The metallicity of the ISM $Z$ evolves with galaxy age $t$ and it is calculated from \citet{Woosley1995} SN II models.
Since only the evolutionary track table with metallicities $Z = (0.005, 0.02, 0.2, 0.4, 1,0, 2.5, 5.0)~Z_\odot$ is prepared, they use the interpolated value.
\Pegase\ assumes that a star releases metal into ISM only at the end of its life, and the recycling model is not instantaneous.
The library of stellar spectra used by \Pegase.2 is divided into two according to effective temperature $T_\mr{eff}$.
For $T_\mr{eff} < 50000~\mr{K}$, the library comes from \citet[corrected version (BaSeL-2.0)]{Lejeune1997, Lejeune1998}.

A monochromatic luminosity from total stars at time $t$ is calculated by weighting $L_\lambda^{\mr{SSP}}$ at galaxy age $t^\prime$ with star formation rate SFR,
\begin{equation}
	L_\lambda(t) = \int^{t^\prime = t}_{t^\prime = 0}
	\int^{Z = Z_\mr{max}(t - t^\prime)}_{Z = 0}
	\mr{SFR}(t - t^\prime)
	L_\lambda^\mr{SSP}(t^\prime,Z(t - t^\prime))
	~\mr{d}Z \dt^\prime, 
\end{equation}
where $Z_\mr{max}(t - t^\prime)$ is the maximum metallicity at time $t - t^\prime$.
In order to take into account the stars that were born at the time of the galaxy's birth to the stars that are just born, the value is integrated over time.
The time lag $t - t^\prime$ represents the time difference between the formation of a star and the end of the evolution of the star.
We chose the Schmidt law \citep[][Equation (\ref{equ:SFR})]{Schmidt1959} for the SFR. 

\subsection {Dust properties} \label{sec:Dust_properties}
Radiative transfer is the method to calculate the propagation of energy in systems of various sizes (from isolated gas clouds to galaxies).
In the galaxy ISM, radiation is mainly affected by absorption and scattering by dust grains.
One of the easiest ways to calculate radiative transfer is to assume that the ISM has a homogeneous distribution.
However, it has been observed that actual galaxies have a more complex structure in general \citep[e.g.,][]{Field1969, McKee1977}.
If a homogeneous dust distribution is assumed, the optical depth of the dust is larger than that in the case of inhomogeneous distribution.
In other words, if the dust mass is estimated with a homogeneous distribution, the attenuation per dust grain is overestimated, and the dust mass would be underestimated as a consequence.
Therefore, in this paper, we consider a clumpy dust distribution.

The calculation of three-dimensional radiative transfer with clumpy dust usually requires substantial computational cost.
\citet{Neufeld1991} and \citet{Hobson1993} introduced the method that solves the radiative transfer with MGA in a one-dimensional plain parallel galaxy \citep{Varosi1999, Inoue2005}.
The MGA treats the dusty region as a kiloparsec-size huge grain called mega-grain, and regards absorption and scattering behave in the same way as typical grains with effective optical properties.
We approximate the complex distribution of stars, dust grains, and gas in the model galaxy, to simplify costly calculations in three-dimensional space.
In this approximation, the distribution of young stars is clumpy and the young stars are embedded by mega-grain.
In contrast, old stars are supposed to distribute smoothly in a diffuse way.
The light emitted by young stars is stronger attenuated than the light emitted by older stars due to the surrounding mega-grains.
\cite{Inoue2005} researched the effect of changing criterion of young star $t_\mr{y}$.
He conclude 10~Myr is best fit to the MW attenuation, and we apply it in this paper.
Assuming thermal and chemical equilibrium with temperature $T <10^4$~K, the ISM is represented by two phases, WNM and CNM \citep[e.g.,][]{Wolfire2003, Koyama2002}.
The relation of the thermal pressure with hydrogen density is expressed by fitting the phase diagram \citep{Inoue2005} as,
\begin{align}
  \frac{p/k}{10^4~\mr{K\,cm^{-3}}}
  & = \frac{n_{\mr{H, WNM}}}{1~\mr{cm^{-3}}},~& (\mr{WNM}) \label{equ:n_WNM} \\
  \frac{p/k}{10^{4.5}~\mr{K\,cm^{-3}}}
    & = \left( \frac{n_{\mr{H, CNM}}}{10^3~\mr{cm^{-3}}}\right)^{0.7},~& (\mr{CNM}) \label{equ:n_CNM}
\end{align}
where $p$ is the pressure.
$n_\mr{H, WNM}$ and $n_\mr{H, CNM}$ are the hydrogen density of WNM and CNM, respectively.
We regard the WNM as a homogeneous interclump medium and the CNM as a clump.
The clump radius $r_\mr{cl}$ is calculated by assuming it to be self-gravitating \citep{Inoue2005},
\begin{equation}
  r_\mr{cl}
  = \frac{1}{\rho_\mr{cl}}
  \sqrt{\frac{15p}{4\pi G}}
  = \frac{1}{\mu m_\mr{p}n_\mr{H, CNM}}
  \sqrt{\frac{15p}{4\pi G}}
  \sim 10.4~\mr{pc}, \label{equ:r_cl}
\end{equation}
where $\rho_\mr{cl} = \mu m_\mr{p} n_\mr{H, CNM}$ is the clump density, $\mu$ is the mean atomic density, $G$ is the gravitational constant and $m_\mr{p}$ is the proton mass.
Clumps exist in the interclump medium.
We assume all clumps are spherical and have a constant radius and density.
In this approximation, we use the mass absorption and scattering coefficient and the scattering asymmetry parameter of dust grain, $k_\mr{abs}$, $k_\mr{scat}$ and $g_\mr{d}$ respectively, averaged by dust size distribution calculated by the Asano model:
\begin{align}
    k_\mr{abs} 
    & = \frac{\int^{a_\mr{max}}_{a_\mr{min}}
    \pi a^2 Q_\mr{abs}(a)f(a)\da}
    {\int^{a_\mr{max}}_{a_\mr{min}}
    \md(a) f(a) \da}, \\
    k_\mr{scat}         
    & = \frac{\int^{a_\mr{max}}_{a_\mr{min}}
    \pi a^2 Q_\mr{scat}(a) f(a) \da}
    {\int^{a_\mr{max}}_{a_\mr{min}}
    \md(a) f(a) \da}, \\
    g_\mr{d} 
    & = \frac{\int^{a_\mr{max}}_{a_\mr{min}}
    g(a) \pi a^2 Q_\mr{scat}(a) f(a)\da}
    {\int^{a_\mr{max}}_{a_\mr{min}}
    \pi a^2 Q_\mr{scat}(a) f(a) \da},
\end{align}
where $f(a)$ is the dust number distribution, $Q_\mr{abs}(a)$ and $Q_\mr{scat}(a)$ are the absorption and scattering coefficient, and $g(a)$ is the scattering asymmetry parameter of a grain, respectively.
In this model $Q_\mr{abs}(a)$, $Q_\mr{scat}(a)$, and $g(a)$ are calculated by the Mie theory \citep{Bohren1983}.
We adopt \cite{Draine1984} and \cite{Laor1993} for silicate and graphite, and \cite{Li2001} for PAH as optical parameters.
The mass extinction coefficient and scattering albedo are defined as
\begin{equation}
    k_\mr{d} = k_\mr{abs} + k_\mr{scat}.
\end{equation}
In the MGA, we replace the optical properties, namely the extinction coefficient per unit length $\kappa$, the scattering albedo $\omega$, and the scattering asymmetry parameter $g$ with the effective ones.
The relative optical depth of a clump with the interclump medium is
\begin{equation}
  \tau_\mr{cl} = (\rho_\mr{cl}
  - \rho_\mr{icm})k_\mr{d}Dr_\mr{cl}.
\end{equation}
where $D$ is the dust-to-gas mass ratio calculated by the Asano model.
The extinction coefficient per unit length of the medium by clump is 
\begin{equation}
  \kappa_\mr{mg}  
  = n_\mr{cl}\pi r^2_\mr{cl}
  P_\mr{int}(\tau_\mr{cl}) = \frac{3f_\mr{cl}}{4r_\mr{cl}}P_\mr{int}(\tau_\mr{cl}),
\end{equation}
where $n_\mr{cl}$ is the number density of clump and $f_\mr{cl}$ is the clump filling fraction,
\begin{equation}
    f_\mr{cl} = \frac{n_\mr{H} - n_\mr{H,WNM}}{n_\mr{H,CNM} - n_\mr{H,WNM}} \label{equ:f_cl}.
\end{equation}
We assume that mean hydrogen number density in the galaxy $n_\mr{H}$ has a constant value 1~cm$^3$.
$P_\mr{int}(\tau)$ is the interaction probability against parallel light by a sphere with optical depth $\tau$, and represented as
\begin{equation}
  P_\mr{int}(\tau) = 1 - \frac{1}{2\tau^2}
  + \left( \frac{1}{\tau} + \frac{1}{2\tau^2}
  \right)e^{-2\tau}.
\end{equation}
This equation is obtained by integrating the light incident on the sphere in all directions and taking the ratio to the case where the optical depth of the sphere is zero (see appendix C of \citet{Varosi1999} for details).
The extinction coefficient of interclump medium is
\begin{equation}
  \kappa_\mr{icm} = k_\mr{d}D\rho_\mr{icm}.
\end{equation}
Thus, the effective extinction coefficient is expressed as
\begin{equation}
  \kappa_\mr{eff} = \kappa_\mr{mg} + \kappa_\mr{icm}.
\end{equation}
The scattering albedo of clump is
\begin{equation}
  \omega_\mr{cl} = \omega_\mr{d}P_\mr{esc}
  (\tau_\mr{cl}, \omega_\mr{d}),
\end{equation}
where $\omega_\mr{d} = k_\mr{scat}/k_\mr{d}$ is the scattering albedo of normal grain averaged by grain size distribution and
\begin{equation}
  P_\mr{esc}(\tau, \omega)
  = \frac{\frac{3}{4\tau}P_\mr{int}(\tau)}{1 - \omega \left[ 1 - \frac{3}{4\tau}P_\mr{int}(\tau)\right]}, \label{equ:escape_probability}
\end{equation}
is the photon escape probability from a sphere grain.
The effective scattering albedo is
\begin{equation}
  \omega_\mr{eff}
  = \frac{\omega_\mr{cl} \kappa_\mr{mg}
  + \omega_\mr{d}\kappa_\mr{icm}}{\kappa_\mr{eff}}.
\end{equation}

The light entering the clump is scattered by the dust in the clump and escapes in various directions.
An optical parameter that indicates in which direction light escapes is called an asymmetry parameter of clump and it is defined as $g_\mr{cl} = \langle \cos\theta_\mr{esc} \rangle$, where $\theta_\mr{esc}$ is the angle between enter and escape directions.
$g_\mr{cl}$ is given by fitting the Monte Carlo calculation result in \citet{Varosi1999}, and represented by the following empirical formula,
\begin{equation}
  g_\mr{cl}(\tau_\mr{cl}, \omega_\mr{cl}, g_\mr{d})
  = g_\mr{d} - C
  \left(
  1 - \frac{1 + e^{-B/A}}{1 + e^{(\tau_\mr{cl} - B)/A}}
  \right),
\end{equation}
where
\begin{align}
  A & \equiv 1.5 + 4g_\mr{d}^3 + 2\omega_\mr{d}\sqrt{g_\mr{d}}
  \exp (-5g_\mr{d}), \\
  B & \equiv 2 - g_\mr{d}(1 - g_\mr{d}) - 2 \omega_\mr{d}g_\mr{d}, \\
  C & \equiv \frac{1}{3 - \sqrt{2g_\mr{d}} - 2\omega_\mr{d}g_\mr{d}(1 - g_\mr{d})}.
\end{align}
The effective asymmetry parameter is
\begin{equation}
  g_\mr{eff} =
  \frac{g_\mr{cl}\kappa_\mr{mg} + g_\mr{d}\kappa_\mr{icm}}{\kappa_\mr{eff}}.
\end{equation}

\subsection{Radiative transfer in a one-dimensional galaxy}
\label{sec:Radiative_transfer_in_one_dimensional_galaxy}

\begin{figure}
  \includegraphics[width=\columnwidth]{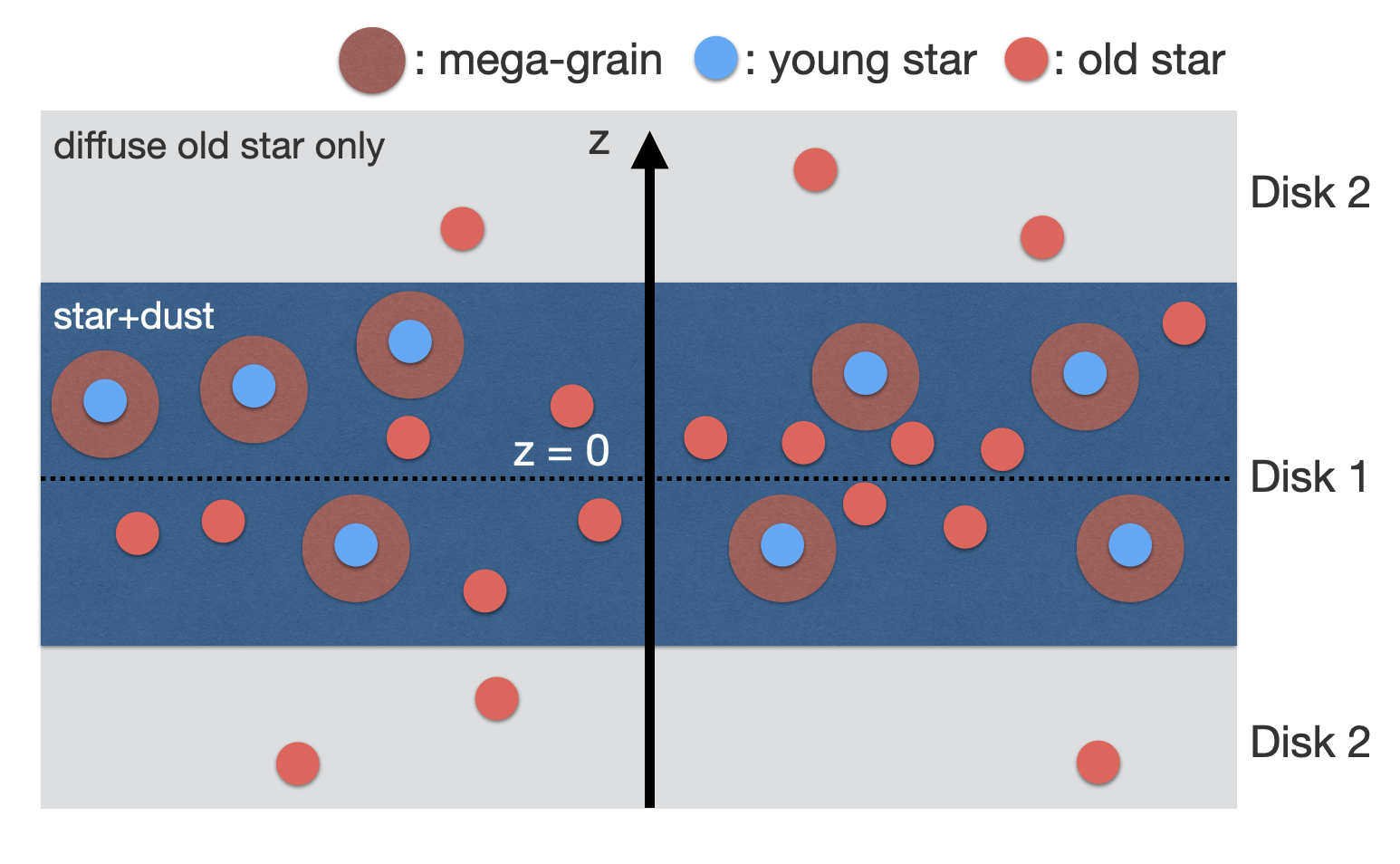}
  \caption{The geometry of a one-dimensional plain parallel galaxy.}
  \label{fig:one_dimensional_galaxy}
\end{figure}
We assume a one-dimensional plane-parallel galaxy along the $z$-axis for solving radiative transfer shown in Figure~\ref{fig:one_dimensional_galaxy}.
We set two kinds of disks in the model.
One is a gas+dust disk containing young stars (disk~1).
Stars are born in cold, dense regions (CNM), so we assume young stars are surrounded by clamps, and disk~1 is also full of interclump medium.
Disk~1 has a constant density of stars, gases, and grains. Optical depth $\tau$ is defined with constant effective extinction coefficient $\kappa_\mathrm{eff}$ as
\begin{equation}
  \mathrm{d}\tau = -\kappa_\mathrm{eff}\,\mathrm{d}z,
\end{equation}
with $\tau = 0$ at $z = h_\mr{d}$ and $\tau = \kappa_\mr{eff}h_\mr{d}$ at $z = 0$.
Where $2h_\mr{d}$ is the thickness of disk~1.
The other disk contains only exponentially and smoothly distributed old stars (disk~2).
The thickness of the disk~2 is $4h_\mr{d}$, which is twice larger than that of  disk~1.
These two disks are stacked so that their centers are aligned.
In the above condition, radiative transfer is formulated as,
\begin{equation}
  \mu \diff{I(\tau, \mu)}{\tau}
  = -I(\tau, \mu) + S(\tau, \mu), \label{equ:radiative_transfer}
\end{equation}
where $I(\tau, \mu)$ is the specific intensity at $\tau$ and $\mu \equiv 1/\cos \theta$. 
The $\theta$ is the angle between the ray and $z$-axis. 
Source function $S$ is represented as
\begin{equation}
  S(\tau, \mu) = \frac{\eta_\ast(\tau)}{\kappa_\mr{eff}}
  + \omega_\mr{eff}
  \int^1_{-1}
  I(\tau, \mu^\prime)
  \Phi(g_\mr{eff}, \mu, \mu^\prime)\mr{d}\mu^\prime, \label{equ:source_function}
\end{equation}
where $\eta_\ast$ is the stellar emissivity and $\Phi$ is the scattering phase function. 
Here we adopt the Henyey-Greenstein phase function \citep{Henyey_Greenstein1941}

The first term in the right hand side of Equation (\ref{equ:source_function}) represents the intensity of light emitted from the star that escapes from the clump where the star was born.
The second term in the right hand side represents the integral of the light scattered in the considering direction $\mu$ among the scattered light by dust grain.
The boundary conditions in this galaxy at $z = 0$ and $z = h_\mr{d}$ are
\begin{align}
    I(\tau = \kappa_\mr{eff}h_\mr{d},\mu)
    & = I(\tau=\kappa_\mr{eff}h_\mr{d},-\mu), \\
    I(\tau = 0,\mu < 0) & = -\frac{\int^\infty_{h_\mr{d}}\eta_\ast(z)\,\mr{d}z}{\mu}.
\end{align}

The stellar emissivity is normalized by
\begin{equation}
  \int^\infty_{-\infty}\eta_\ast(z)\mr{d}z = 1. \label{equ:emissivity_normalization}
\end{equation}
The intrinsic emissivity from young stars in disk 1 $\eta^\mr{young}_\mr{\ast}$is $1/2h_\mr{d}$ for $|z| \leq h_\mr{d}$ and zero for $|z| > h_\mr{d}$ because it is normalized by Equation (\ref{equ:emissivity_normalization}).
The energy emitted by young star is absorbed by the dust in clump surrounding the star.
Since the escape probability from the clump is Equation (\ref{equ:escape_probability}), the emissivity from young stars is represented as
\begin{equation}
  \eta^\mr{young}_\mr{\ast}
  \begin{cases}
    P_\mr{est}(\tau_\mr{cl}, \omega_\mr{cl})/2h_\mr{d} & (|z| \leq h_\mr{d}) \\
    0 & (|z| > h_\mr{d})
  \end{cases}.
\end{equation}
The old stars in disk 2 are distributed with exponential diffusion along the $z$-axis.
From the normalization Equation $(\ref{equ:emissivity_normalization})$, the emissivity from the old stars is 
\begin{equation}
    \eta_\mr{\ast}^\mr{old}(z) = \frac{e^{-|z|/2h_\mr{d}}}{4h_\mr{d}}.
\end{equation}
The total stellar emissivity at $z$ is represented as
\begin{equation}
    \eta_\ast(z) = f_\mr{y}(t)\eta_\mr{\ast}^\mr{young}(z) + (1 - f_\mr{y}(t))\eta_\mr{\ast}^\mr{old}(z),
\end{equation}
where $f_\mr{y}(t)$ is the luminosity fraction emitted by young stellar at age $t$, it is calculated by
\begin{equation}
    f_\mr{y}(t) = \frac{\int^{\mr{min}[t_\mr{y},t]}_{0}
    \int^{\mr{Z}_\mr{max}(t-t^\prime)}_{0} \mr{SFR}(t-t^\prime) L_\lambda^\mr{SSP}(t^\prime,\mr{Z}(t-t^\prime))\,\mr{dZ}\,\mr{d}t^\prime}
{L_\lambda(t)}.
\end{equation}

In the radiative transfer calculation, we calculate Equation (\ref{equ:radiative_transfer}) and (\ref{equ:source_function}) iteratively until the ratio to the previous loop of the source function in all directions on the galaxy surface converged to 10$^{-10}$.

\subsection{Dust temperature distribution} \label{sec:Dust_temperature_distribution}

The UV and optical photons emitted by stars heat dust grains.
The heated dust grains release the energy by emission from MIR to FIR wavelength photons. 
Large grains have an equilibrium temperature determined by the stellar radiation field.
In contrast, very small grains cannot establish radiative equilibrium, and it does not have a stable equilibrium temperature \citep{Draine1985, Draine2001, Li2001, Takeuchi2003, Takeuchi2005, Horn2007}.
Since they have small heat capacities, they are easily heated by photons and then rapidly cooled
Thus, the (instantaneous) temperature of very small grains is inevitably stochastic, we calculate temperature distribution by Monte Carlo simulation.

\subsubsection{Stochastic heating} \label{sec:Stochastic_heating}

The rate at which a dust grain absorbs photons in the energy range $[E, E + \mr{d}E]$ and time interval $[t, t + \dt]$ is expressed as
\begin{equation}
  \mr{d}p(a, \lambda) = \pi a^2 Q_\mr{abs}(a, \lambda)
  \bar{u}_\lambda \frac{\lambda^3}{h_\mr{p}^2c} \mr{d}E \dt, \label{equ:dp}
\end{equation}
where $\bar{u}_\lambda$ is the mean energy density per wavelength in a galaxy, $h_\mr{p}$ is the Plank constant, and $c$ is the speed of light, respectively \citep[e.g.,][]{Draine1985,Takeuchi2003,Takeuchi2005}. 
The energy density actually varies depending strongly on its spatial position in a galaxy.
Therefore, considering the different energy densities for each position, the calculation time becomes enormous.
Thus, we use the mean energy density $\bar{u}_\lambda$ which is calculated in the same way as \cite{Fioc2019}:
\begin{equation}
L_\lambda^0 - L_\mr{obs}
= c\bar{u}_\lambda k_\mr{abs} M_\mr{d},
\end{equation}
where $L_\lambda^0$ and $L_\mr{obs}$ are the intrinsic stellar luminosity and observed luminosity calculated by transfer radiation.
Therefore, the mean energy density is represented as
\begin{equation}
    \bar{u}_\lambda = \frac{L_\lambda^0 - L_\mr{obs}}{ck_\mr{abs}M_\mr{d}}.
\end{equation}

Equation (\ref{equ:dp}) can be regarded as the probability density distribution if the time interval $\dt$ is appropriately small.
For each dust size, $\dt$ is determined so that the maximum collision probability among all wavelengths is 0.01,
\begin{equation}
 \dt(a) = 0.01 \left[ \pi a^2 Q_\mr{abs}(a, \lambda)
  \bar{u}_\lambda \frac{\lambda^3}{h_\mr{p}^2c} \mr{d}E \right]^{-1}.
\end{equation}

For simplicity, we assume that the energy of an absorbed photon is totally used to heat the dust grains, represented as
\begin{equation}
  E(T + \Delta T) = E(T) + \frac{h_\mr{p}c}{\lambda},
\end{equation}
where $E(T)$ is the enthalpy of dust grains at temperature $T$ and $\Delta T$ is the increment of temperature.
We adopt the Debye model for calculating the enthalpy of dust grains \citep{Li2001}.
The enthalpy of silicate and graphite grains are
\begin{align}
  E_\mr{sil}(T) & = (N_\mr{atom} - 2)k
  \left[
  2f_2 \left( \frac{T}{500~\mr{K}} \right) + f_3 \left( \frac{T}{1500~\mr{K}}
  \right)
  \right], \\
  E_\mr{gra}(T) & = (N_\mr{C} - 2)k
  \left[
  f_2 \left( \frac{T}{863~\mr{K}} \right) + 2f_2 \left( \frac{T}{2504~\mr{K}}
  \right)
  \right], \label{equ:C_gra}
\end{align}
where
\begin{equation}
  f_n(a) \equiv n \int^1_0 \frac{y^n\mr{d}y}{\exp (y/x) - 1}. \label{equ:debye}
\end{equation}
The subscripts 'sil' and 'gra' represent the silicate and graphite grains, respectively.
Equation (\ref{equ:debye}) is the $n$ dimensional Debye function. $N_\mr{atom}$ and $N_\mr{C}$ are the number of atoms in a grain, they are expressed as,
\begin{equation}
  N = \dfrac{\frac{4}{3}\pi a^3 \rho N_\mr{A}}{M},
\end{equation}
where $\rho$ is the mass density, $M$ is the mass number and $N_\mr{A}$ is the Avogadro constant.
For carbonaceous (graphite or PAH) grain, $\rho = 2.26~\mr{g/cm^3}$ and $M = 12.0~\mr{g/mol}$\citep{Draine1984}.
For silicate grain, $\rho = 3.50~\mr{g/cm^3}$ and $M = 172.25~\mr{g/mol}$\citep{Li2001}.
In polycyclic aromatic hydrocarbon (PAH) grains, we consider C-C bond modes are same as graphite and C-H bond modes component is added to Equation (\ref{equ:C_gra}),
\begin{equation}
  E_\mr{pah}(T) = E_\mr{gra} + \frac{\mr{H}}{\mr{C}} N_\mr{C}
  \sum^3_{j=1} \left( \frac{h_\mr{p}\nu_j}{\exp(h_\mr{p}\nu_j/kT) - 1}
   \right). \label{equ:Epah}
\end{equation}
The index $j$ represents the C-H out-of-plane bending modes ($\nu_1 / c = 886~\mr{cm^{-1}}$), in-plane bending modes ($\nu_2/c = 1161~\mr{cm^{-1}}$), and stretching modes ($\nu_3/c = 3030~\mr{cm^{-1}}$), respectively \citep{Draine2001}.
$\frac{\mr{H}}{\mr{C}}$ is the hydrogen to carbon ratio.
We adopt the following empirical formula \citep{Li2001},
\begin{equation}
  \frac{\mr{H}}{\mr{C}} =
  \begin{cases}
    0.5 & (N_\mr{C} < 25) \\
    \frac{0.5}{\sqrt{N_\mr{C}/25}} & (25 < N_\mr{C} < 100) \\
    0.25 & (N_\mr{C} > 100)
  \end{cases}.
\end{equation}

\subsubsection{Dust cooling} \label{sec:Dust_cooling}

The equation of emission of dust grains with radius $a$ is formulated as,
\begin{equation}
  4\pi \epsilon(T, a) = 4 \pi \left(\pi a^2\right) \int Q_\mr{abs}(\lambda)
  \frac{2h_\mr{p}c^2}{\lambda^5}
  \frac{\mr{d}\lambda}{\exp\left( \frac{h_\mr{p}c}{\lambda kT}\right) - 1},
\end{equation}
where $T$ is the temperature of dust grain and $\epsilon(T, a)$ is the emission power per unit time per unit solid angle. 
In the case that dust grains do not absorb energy while cooling, since emission energy and changes in internal energy are balanced, the following equation holds,
\begin{equation}
    \frac{\mr{d}E(T,a)}{\mr{d}T}
    \frac{\mr{d}T}{\mr{d}t}
    = -4\pi \epsilon(T,a).
\end{equation}
This equation can not be solved analytically, but we can calculate, but we can get the temperature variation by numerical calculation.

\subsubsection{Result of dust temperature distribution}
\label{sec:Result_of_dust_temperature_distribution}

\begin{figure}
    \includegraphics[width=\columnwidth]{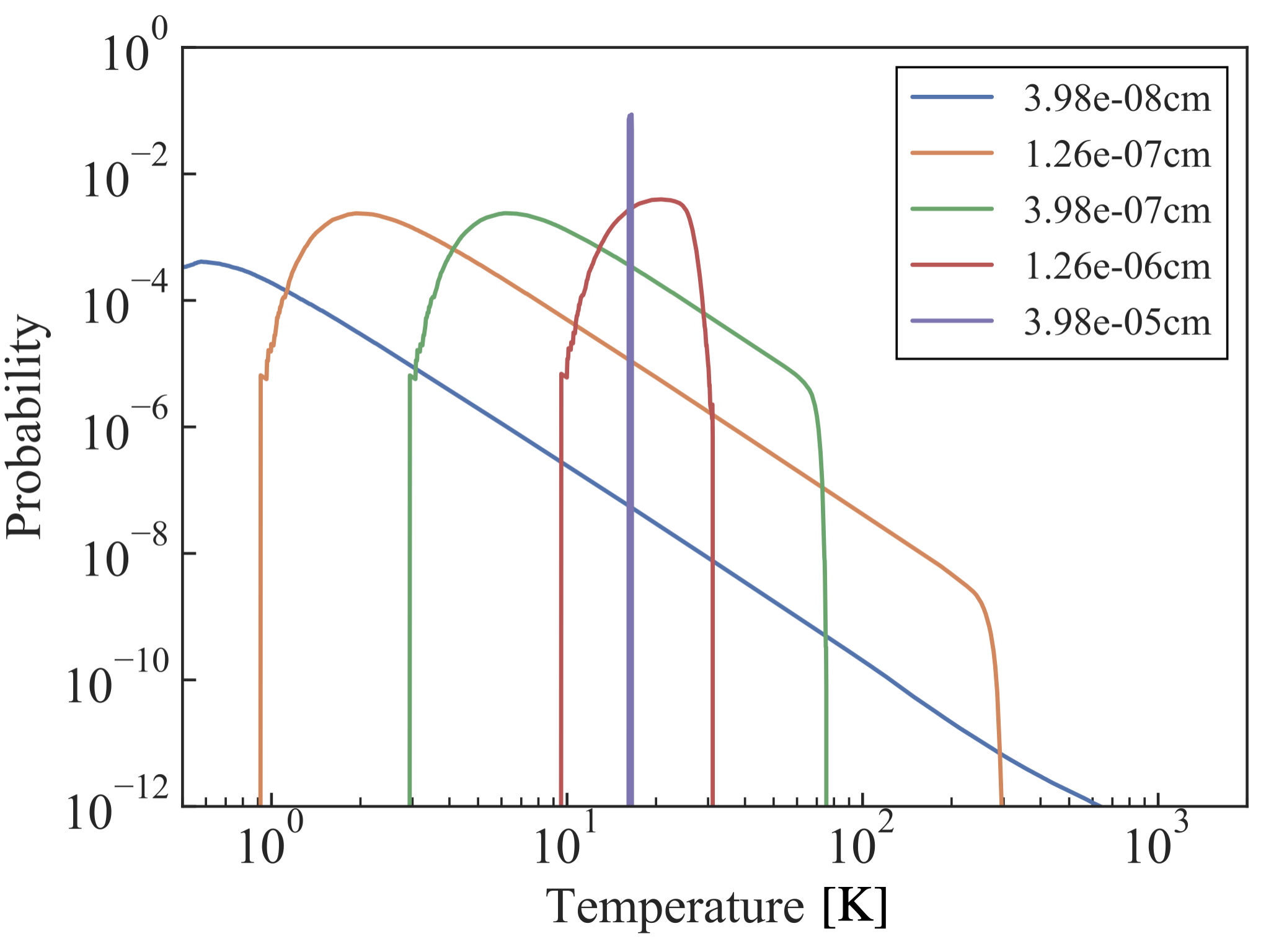}
    \caption{Temperature distribution of the several grain sizes of silicate calculated by Monte Carlo calculation.
    The blue, orange, green, red, and purple curves indicate $3.98 \times 10^{-8}$, $1.26 \times 10^{-7}$, $3.98 \times 10^{-7}$, $1.26 \times 10^{-6}$, and $3.98 \times 10^{-5}$~cm grains, respectively.
    }
    \label{fig:temperature_distribution_Sil}
\end{figure}

Figure~\ref{fig:temperature_distribution_Sil} is the result of dust temperature distribution of silicate.
The galaxy is the face on ($\mu = 1$) MW-like galaxy model at the age of 13~Gyr.
The condition of the galaxy is the same as \S\ref{sec:Result_of_dust_evolution_model}.
The radius of galaxy $R_\mr{gal}$ is 10~kpc and scale height of dust $h_\mr{d}$ is 150~pc which is the typical scale height of cold dust in the MW \citep[e.g.,][]{Binney1998}.

The temperature of small grains is very widely distributed from 1 to 4,000~K.
When grain size becomes larger, the temperature range becomes narrower and approaches the equilibrium temperature.
The equilibrium temperature is represented by \citet{Draine1984, Takeuchi2003} as
\begin{equation}
    T_\mr{eq} \simeq \left( \frac{h_\mr{p}c}{\pi k} \right)
    \left[
    \frac{945u}{960\pi(2\pi A a)h_\mr{p}c}
    \right]^{1/6} \label{equ:Teq},
\end{equation}
and 
\begin{equation}
    u \equiv \int^\infty_0 u_\lambda \mr{d}\lambda,
\end{equation}
where $h_\mr{p}$ is the Planck constant, and we adopt $A_\mr{sil} = 1.34 \times 10^{-3}$~cm for silicate grains \citep{Drapatz1977} and $A_\mr{C} = 3.20 \times 10^{-3}$~cm for carbonaceous grains  \citep{Draine1984}.
When grain size is $3.98 \times 10^{-5}$~cm, the equilibrium temperature of that grain is about 19~K by Equation~(\ref{equ:Teq}) and it is equal to the result calculated by the Monte Carlo calculation.

\begin{figure}
    \includegraphics[width=\columnwidth]{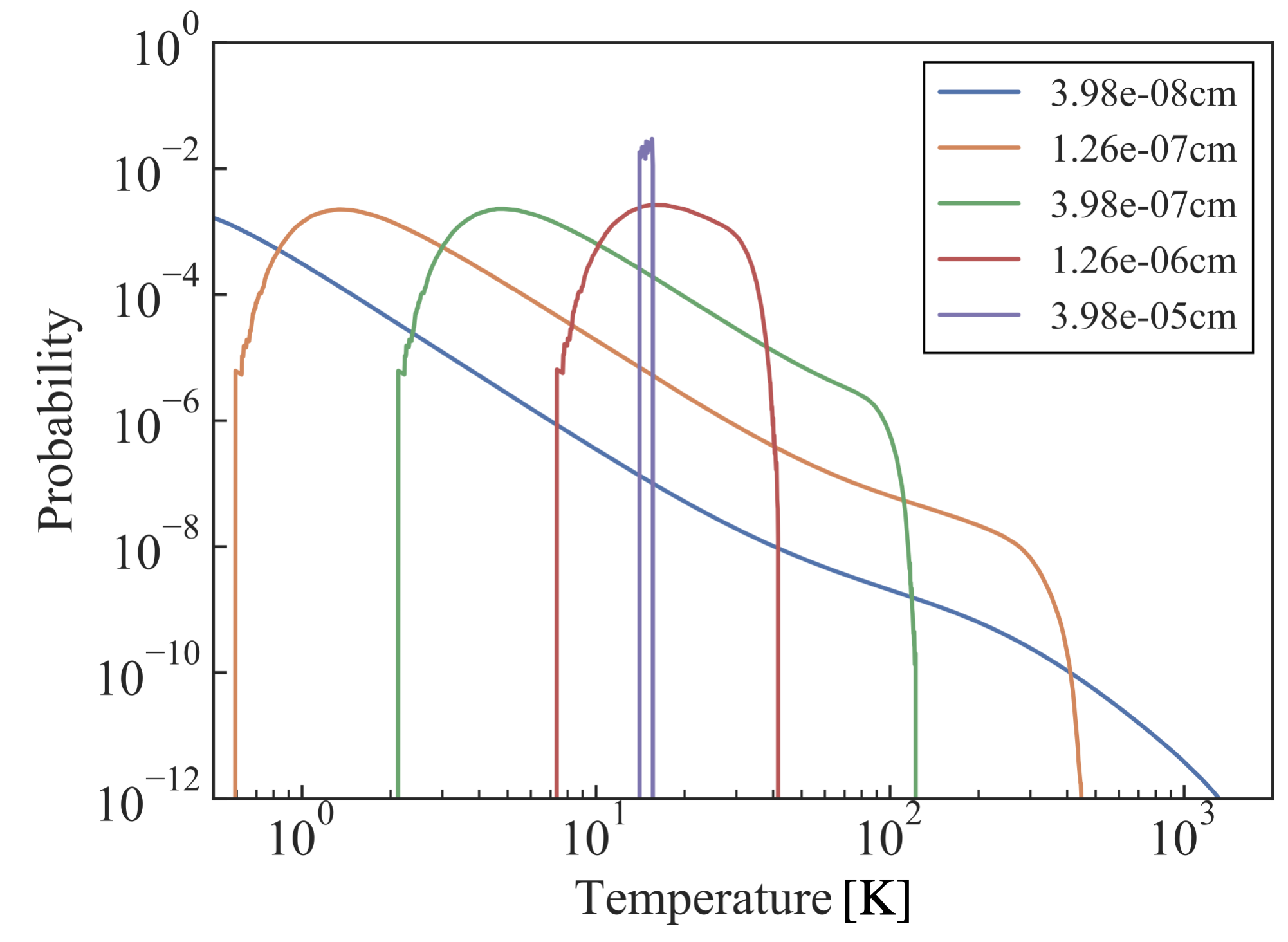}
    \caption{Temperature distribution of the several grain sizes of graphite grain.
    Calculation parameters and color coordinates are the same as Figure \ref{fig:temperature_distribution_Sil}.
    }
    \label{fig:temperature_distribution_Gra}
\end{figure}

Figure~\ref{fig:temperature_distribution_Gra} is the temperature distribution of graphite.
Graphite grains have broad temperature compared with silicate grains in Figure \ref{fig:temperature_distribution_Sil}.
The difference comes from differences of internal energy between silicate and graphite.

\begin{figure}
    \includegraphics[width=\columnwidth]{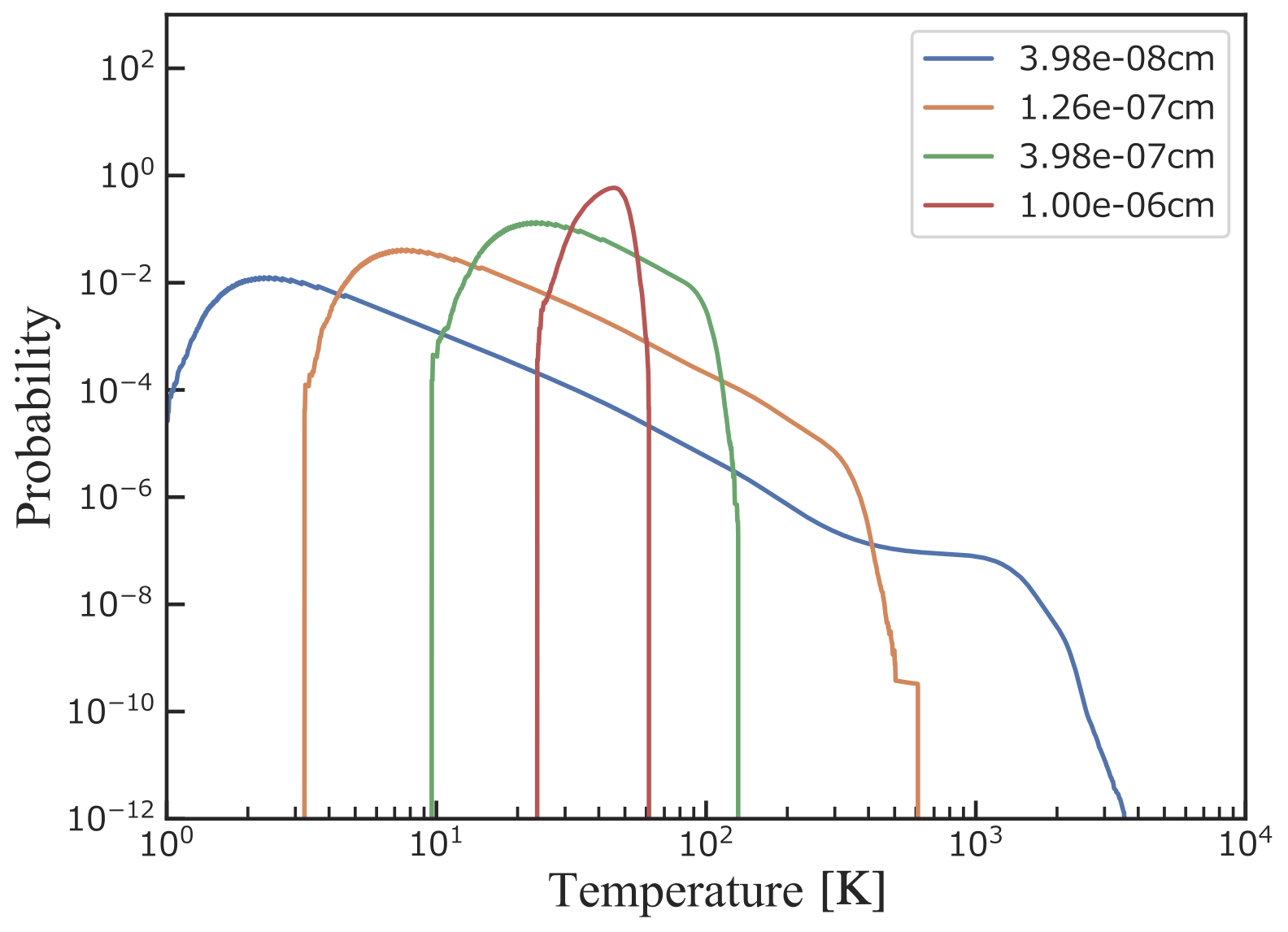}
    \caption{Temperature distribution of the several grain sizes of PAHs.
    Calculation parameters are the same as Figure~\ref{fig:temperature_distribution_Sil}.
    The blue, orange, green, and red curves indicate $3.98 \times 10^{-8}$, $1.26 \times 10^{-7}$, $3.98 \times 10^{-7}$, $1.00 \times 10^{-6}$~cm grains, respectively.
    }
    \label{fig:temperature_distribution_PAH}
\end{figure}

We show the temperature distribution of PAH in Figure~\ref{fig:temperature_distribution_PAH}.
Comparing PAH temperature distribution with graphite, PAHs stay in a narrower temperature range because PAHs have an additional term in the equation of internal energy (Equation~\ref{equ:Epah}).
Almost the same behavior is seen in the result of \citet{Draine2007}.
From the result of temperature distribution, some dust grains might exceed the sublimation temperature, which is 1500~K or higher \citep[e.g.,][]{Baskin2018}.
If we assume that the grain above 1500~K has sublimated and calculated its effect by removing it from the galaxy, the mass of the sublimated grains is only a few percent of the total dust mass.
Thus, the effect of the sublimation for the result is negligible, and we do not consider the effect of sublimation temperature in our model for simplicity.

\subsection{Dust radiation} \label{sec:Dust_radiation}

The dust radiation depends on the temperature distribution $\diff{P_i(a)}{T}$ calculated by the method of the above sections. 
The monochromatic luminosity of a dust grain of species $i$ (silicate, graphite, neutral PAH, or ionized PAH) is expressed as
\begin{equation}
  L_i^{\mr{grain}}(a, \lambda) = 4\pi a^2 \pi \int Q_\mr{abs}^i(\lambda) B_\lambda (T) \diff{P_i(a)}{T}~\mr{d}T, 
\end{equation}
where $B_\lambda$ is the blackbody radiation and $Q_\mr{abs}^i$ is the absorption coefficient of dust species $i$.
Total luminosity at wavelength $\lambda$ is represented as,
\begin{equation}
  L(\lambda) = \sum_i Q_\mr{abs}^i(\lambda)\int L_i^{\mr{grain}}(a, \lambda) f_i(a)~\da.
\end{equation}
$f_i(a)$ is the dust number distribution of dust species $i$ from the dust evolution model.

\section{Result: Milky Way-like galaxy model SED}
\label{sec:Result_Milky_Way_like_galaxy_model_SED}

\begin{figure}
    \includegraphics[width=\columnwidth]{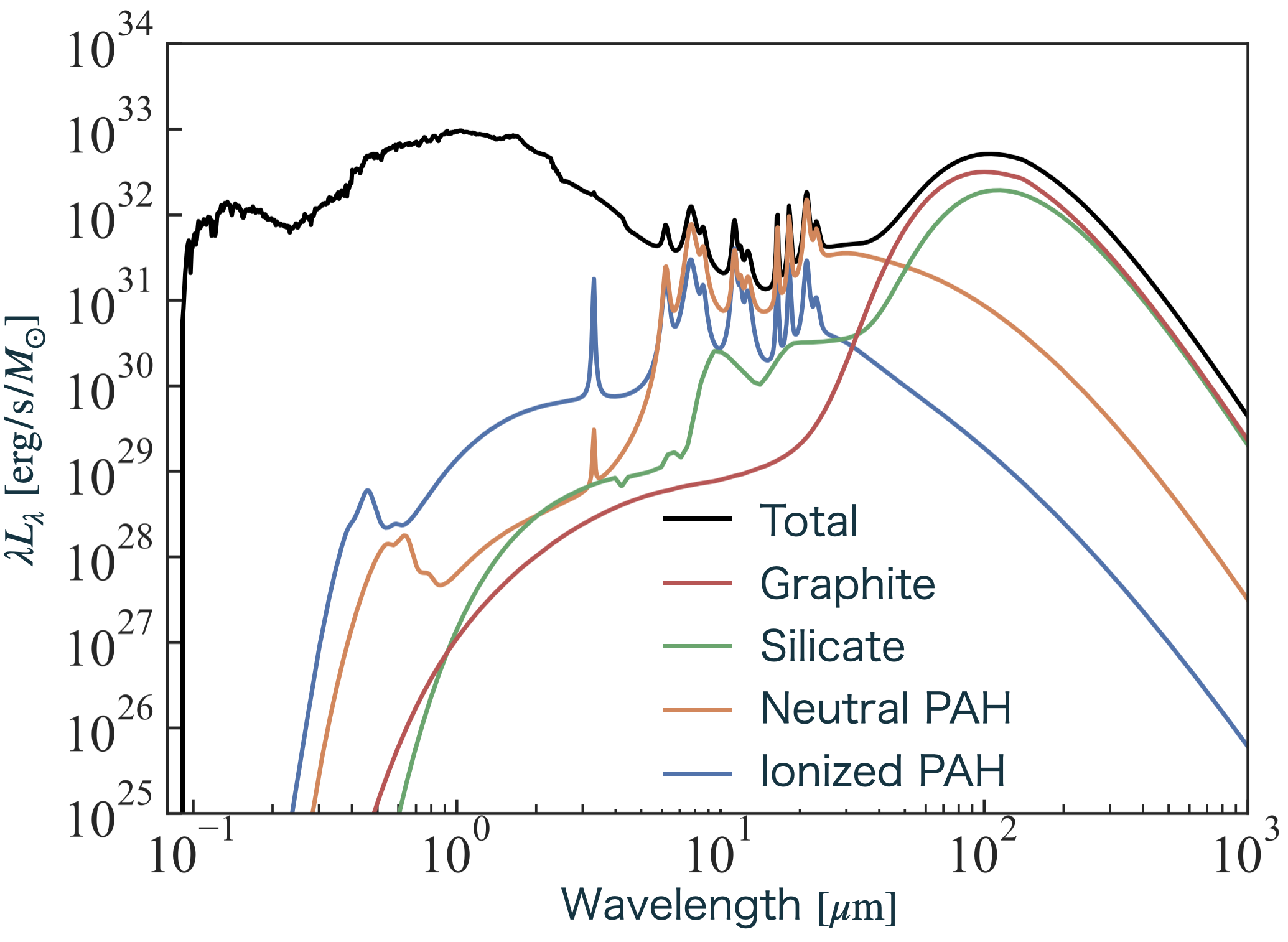}
    \caption{The result of our SED model with MW-like galaxy parameters at an age of 13~Gyr.
    The black curve represents the overall emission of the galaxy.
    Other color curves express each dust species (blue: ionized PAH, orange: neutral PAH, green: silicate, red: graphite).}
    \label{fig:SED_MW_at_13Gyr}
\end{figure}

In Figure \ref{fig:SED_MW_at_13Gyr}, we show the model result with a face-on ($\mu = 1$) MW-like galaxy model (\S\ref{sec:Result_of_dust_temperature_distribution} and \S\ref{sec:Result_of_dust_evolution_model}) at the age of 13~Gyr (the same setting as \S\ref{sec:Result_of_dust_temperature_distribution}).
Each curve in Figure \ref{fig:SED_MW_at_13Gyr} represents the corresponding emission species.
At the 912~$\mathrm{\mathring{A}}$ wavelength, we see the cutoff of the Lyman break.
The UV to IR wavelength region is dominated by stellar emission.
Numerous PAH lines are prominent in the mid-IR, and the far-IR range is dominated by the continuum emission from large graphite grains.
The emission of silicate is weaker than that of graphite in the wavelength range of $200~\mr{\mu m}$ or less, and is effective only in the longer wavelength range.
The temperature of graphite and silicate fitted by a gray body are 28~K and 26~K, respectively.
The difference in emission and temperature between graphite and silicate is caused by the number of grains.

\begin{figure}
    \includegraphics[width=\columnwidth]{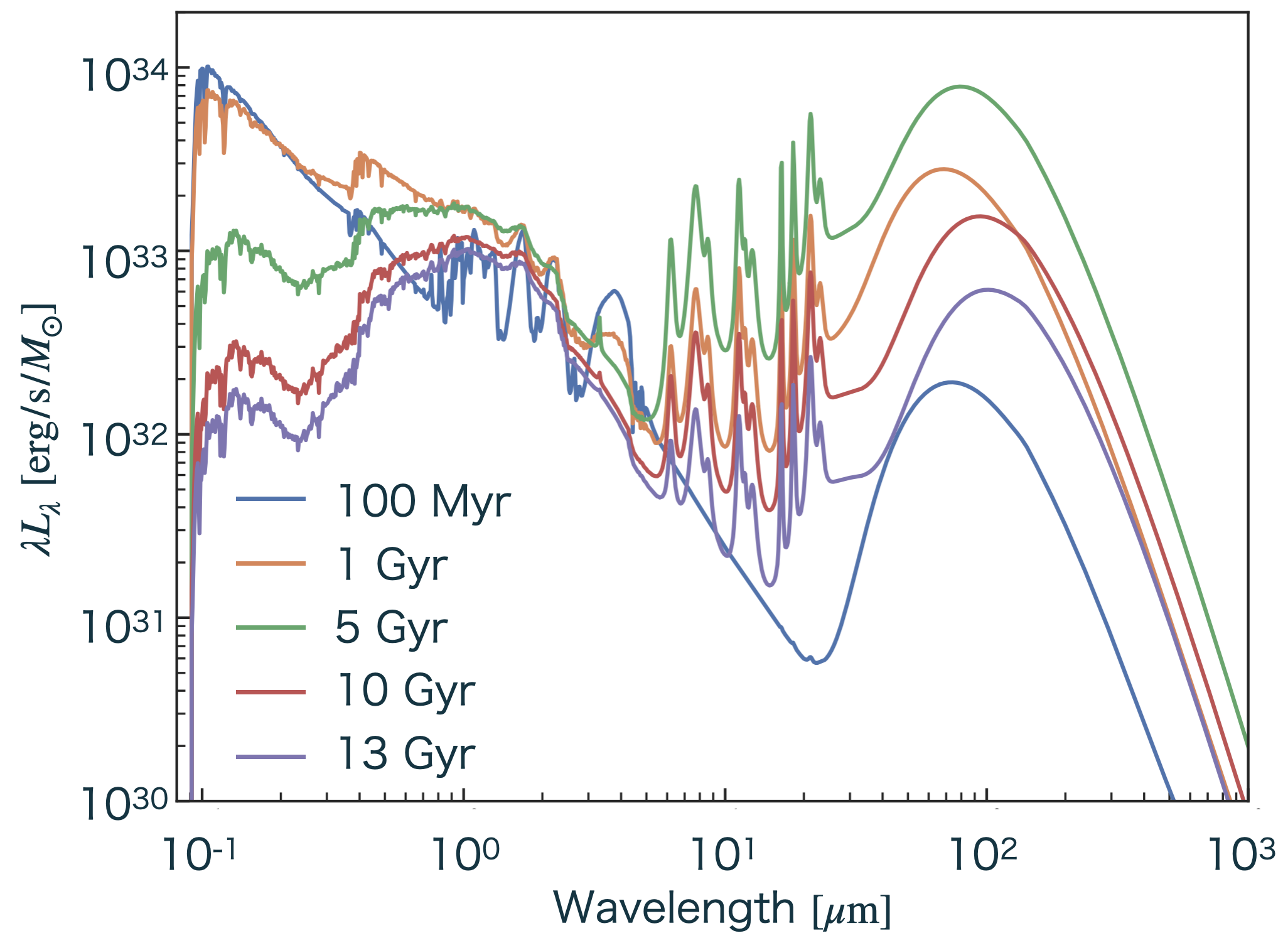}
    \caption{The evolution of the SED of a MW-like galaxy.
    Parameters are the same as Figure~\ref{fig:SED_MW_at_13Gyr}.
    Blue, orange, green, red, and purple curves indicate age of 100~Myr, 1, 5, 10, 13~Gyr, respectively.
    }
    \label{fig:SED_evolution_with_age}
\end{figure}

Figure~\ref{fig:SED_evolution_with_age} is the time evolution of our galaxy SED model with the MW-like galaxy model parameters.
The purple curve represents the SED of a MW-like galaxy at the age of 13 Gyr, as Figure~\ref{fig:SED_MW_at_13Gyr}.
Blue, orange, green, and red curves represent the age of 100~Myr, 1, 5, and 10~Gyr, respectively.
Figure~\ref{fig:SED_evolution_with_age} shows that the UV region emitted by stars monotonically decreases with the evolution.
Since we assume a closed box, the gas mass decreases monotonically as it is consumed by star formation.
The SFR is proportional to the gas mass, hence SFR also decreases monotonically, and the UV radiation also decreases.
The overview of the time evolution of the SED is as follows.
\begin{itemize}
    \item <100~Myr
    \begin{itemize}
        \item Stellar emission dominates the SED and PAH emission does not exist yet.
    \end{itemize}
    \item 100~Myr--1~Gyr
    \begin{itemize}
        \item Stellar emission still dominates the SED, but the dust emission including PAH gradually becomes prominent.
    \end{itemize}
    \item 1--5~Gyr
    \begin{itemize}
        \item Dust emission dominates the SED and dust emission becomes strongest at this age.
    \end{itemize}
    \item 5--13~Gyr
    \begin{itemize}
        \item The emission both from stars and dust gradually decreases, along with the decline of the star formation rate.
    \end{itemize}
\end{itemize}
The details of the SED evolution are explained below.

At 100~Myr, since only a very small amount of dust has been produced, stellar radiation is not attenuated, and dust radiation is weak.
Particularly, PAHs are not produced in young galaxies, hence the MIR radiation is very faint.
For an SED model that supposes a constant size distribution without considering the evolution of the dust size distribution \citep[e.g.,][]{Schurer2009}, many PAHs are observed even in such young galaxies, and different conclusions are deduced.

\begin{figure}
    \includegraphics[width=\columnwidth]{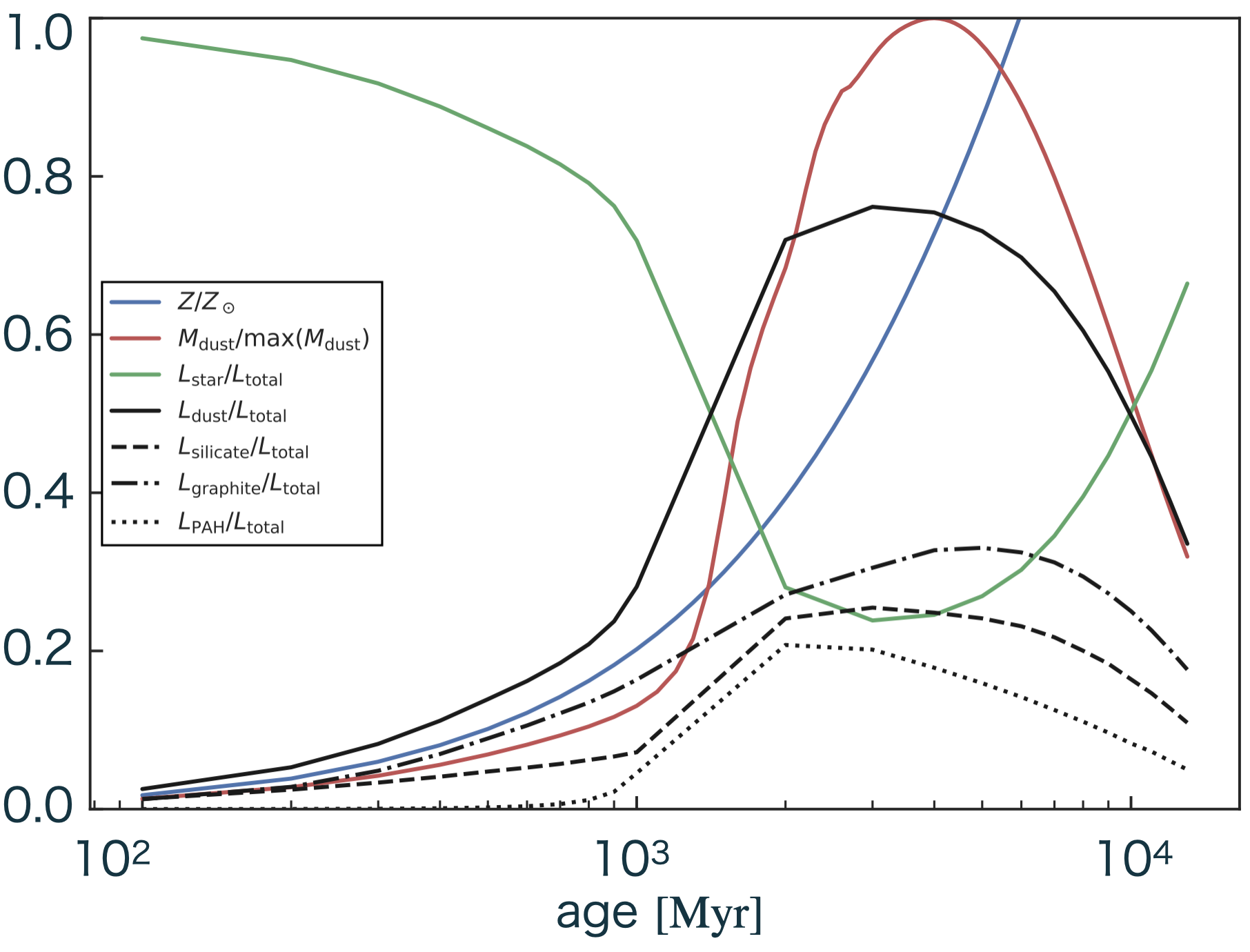}
    \caption{The evolution of metallicity, dust mass, and bolometric luminosity of each component of the galaxy.
    Parameters are set to be the same as Figure~\ref{fig:SED_MW_at_13Gyr}.
    Metallicity, dust mass, and bolometric luminosities are normalized by solar metallicity $Z_\odot = 0.02$ \citep{Anders1989}, maximum value of it, and overall bolometric luminosity, respectively.
    The calculation was performed with the age of the logarithmic scale bin.}
	\label{fig:SED_evolution_with_age_component}
\end{figure}

The evolution of metallicity, dust mass and bolometric luminosity for each component are shown in Figure~\ref{fig:SED_evolution_with_age_component}.
The dust mass is normalized with respect to its maximum value.
Dust mass and luminosity are tightly correlated.
Here we adopt $Z_\odot = 0.02$ \citep{Anders1989}.
At 1~Gyr, the dust mass is gradually increasing, and along with this, the IR radiation from dust becomes prominent. 
Because the PAH mass increases by the evolution in the ISM, their characteristic mid-infrared line emission can be seen.

The dust emission becomes strongest at 3~Gyr if we adopt the star formation timescale $\tau_\mr{SF} = 3$~Gyr.
As predicted, in the MW-like model, when the metallicity exceeds 0.1~$Z_\odot$, the metal accretion process becomes effective and the dust mass increases \citep{Asano2013a}.
Star formation is active, but the UV continuum from young stars is strongly attenuated due to the increase of dust mass.

After 3~Gyr, the dust mass decreases due to the destruction by SN shocks and astration.
Dust radiation also decreases with the age of the galaxy.
This is not only due to the reduction of dust mass, but also due to the decline of the UV light from young stars to heat dust grains.

Focusing on the metallicity, it reaches 1.6~$Z_\odot$ at 13~Gyr, which is larger than the solar metallicity.
This is due to the assumption of the closed box model.
In the closed box model, there is no inflow of gas and outflow of ISM and the metallicity increases monotonically.
However, considering the infall model, the metallicity is reduced because the ISM is diluted by the inflow of gas \citep{Erb2006}.

We should note the difference in the evolution of each species of dust grains in Figure~\ref{fig:SED_evolution_with_age_component}.
The increase of graphite emission is more gradual than the increase of other components.
The metallicity of a galaxy exceeds the critical metallicity in accretion onto the dust surface, leading to the sharp rise of the dust mass and emission  \citep{Asano2013a}.
Shattering produces small dust grains and makes the surface area of dust larger, and consequently leads to the boost of the accretion efficiency.
However, since we regard almost all small carbonaceous grains as PAHs, the mass of graphite grains does not have the discontinuous increase.
The bolometric luminosity of dust emission is dominated by graphite in all epochs.

\section{Discussion}
\label{sec:Discussion_the_effect_of_changing_parameters}

\subsection{Effect of star formation timescale}

\begin{figure*}
    \includegraphics[width=17cm]{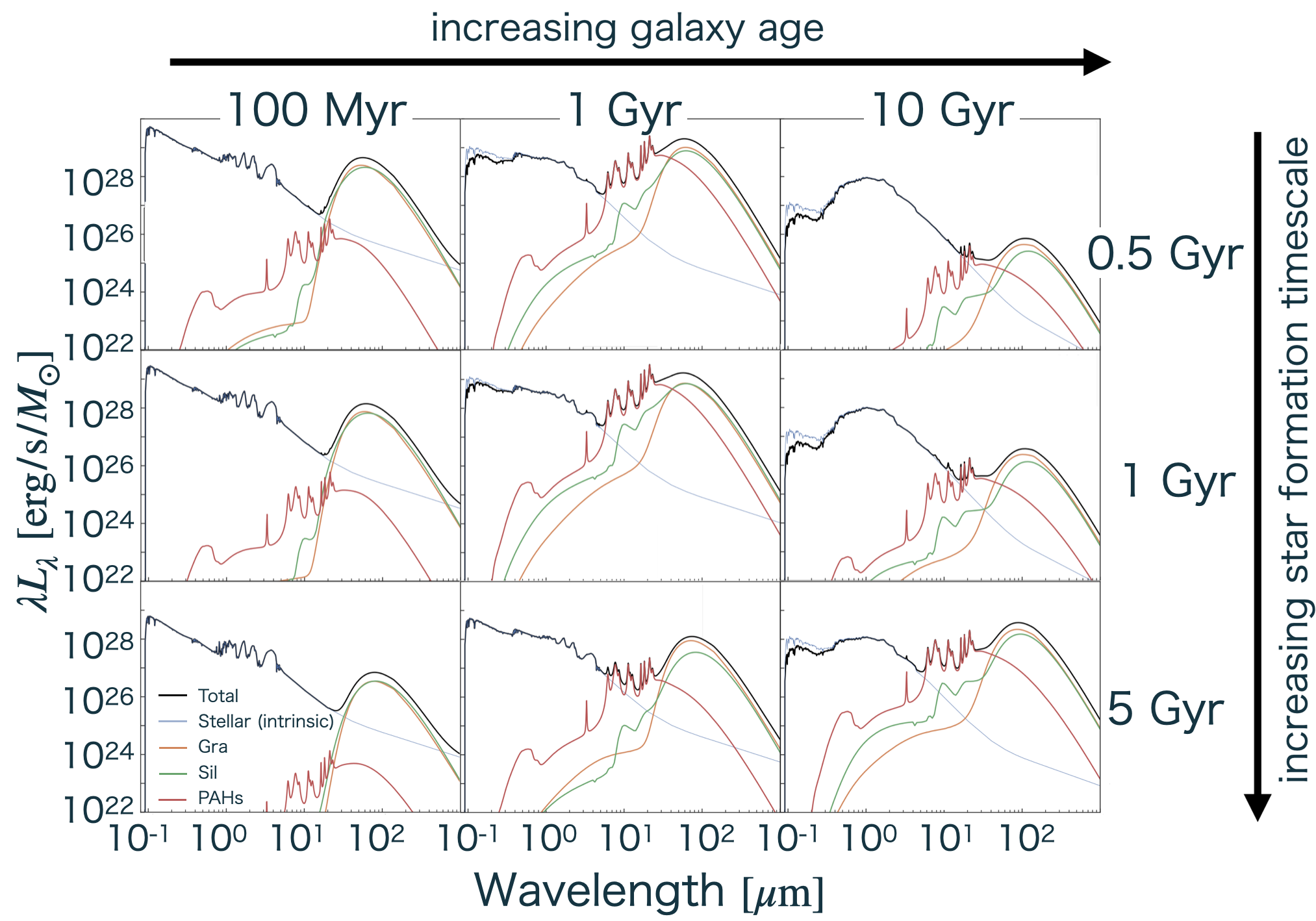}
    \caption{The effect of star formation timescale $\tau_\mr{SF}$ for galaxy SED.
    The geometrical model parameters are the same as Figure \ref{fig:MW_tau_SF}.
    The galaxy age is increasing from left to right, SF timescale increases from top to bottom.
    The age and SF timescale are written on each plot.
    Black, orange, green, red thick curves represent overall,  graphite, silicate, and PAHs luminosity, respectively.
    Blue thin curve is an intrinsic (unattenuated) stellar emission.
    }
	\label{fig:MW_tau_SF}
\end{figure*}

The effect of star formation (SF) timescale $\tau_\mr{SF}$ is shown in Figure~\ref{fig:MW_tau_SF}.
In Figure~\ref{fig:MW_tau_SF}, the SEDs are calculated with different SF timescale and age from Figure \ref{fig:SED_MW_at_13Gyr}, while geometrical parameters are kept the same. 
The galaxy age increase from left to right the panels (100~Myr, 1, and 10~Gyr), and SF timescale increase from top to bottom the panels ($\tau_\mr{SF} = $ 0.5, 1, 5~Gyr).
Three characteristic trends are observed in Figure~\ref{fig:MW_tau_SF}.
First, the UV light emitted by stars and the IR light emitted by graphite and silicate grains at the age of 100~Myr decrease with increasing $\tau_\mr{SF}$.
This is because the age of the galaxy is sufficiently small with respect to the SFR, and the larger the time scale of the star formation rate, the fewer stars form.
In these young galaxies, PAHs are not produced and the PAHs emission is not observed in the model with any  $\tau_\mr{SF}$ yet.
This indicates that the dust mass in early galaxies is dominated by production from stars instead of accretion processes.

Second, the overall bolometric luminosities tend to be stronger when the age of the galaxy is equal to the SF timescale.
In this evolutionary phase, the SFR is still large and a large amount of dust exists in the galaxy.

Lastly, when the age of the galaxy is older than the SF timescale, the galaxy has very weak stellar emission due to consumption of most of the gas in the ISM which is an ingredient of star formation.
The dust emission is also very weak in the galaxy because both, decreasing dust mass and UV light which is the source of heating dust grains.


\subsection{Effect of geometrical parameters}

\begin{figure}
    \includegraphics[width=\columnwidth]{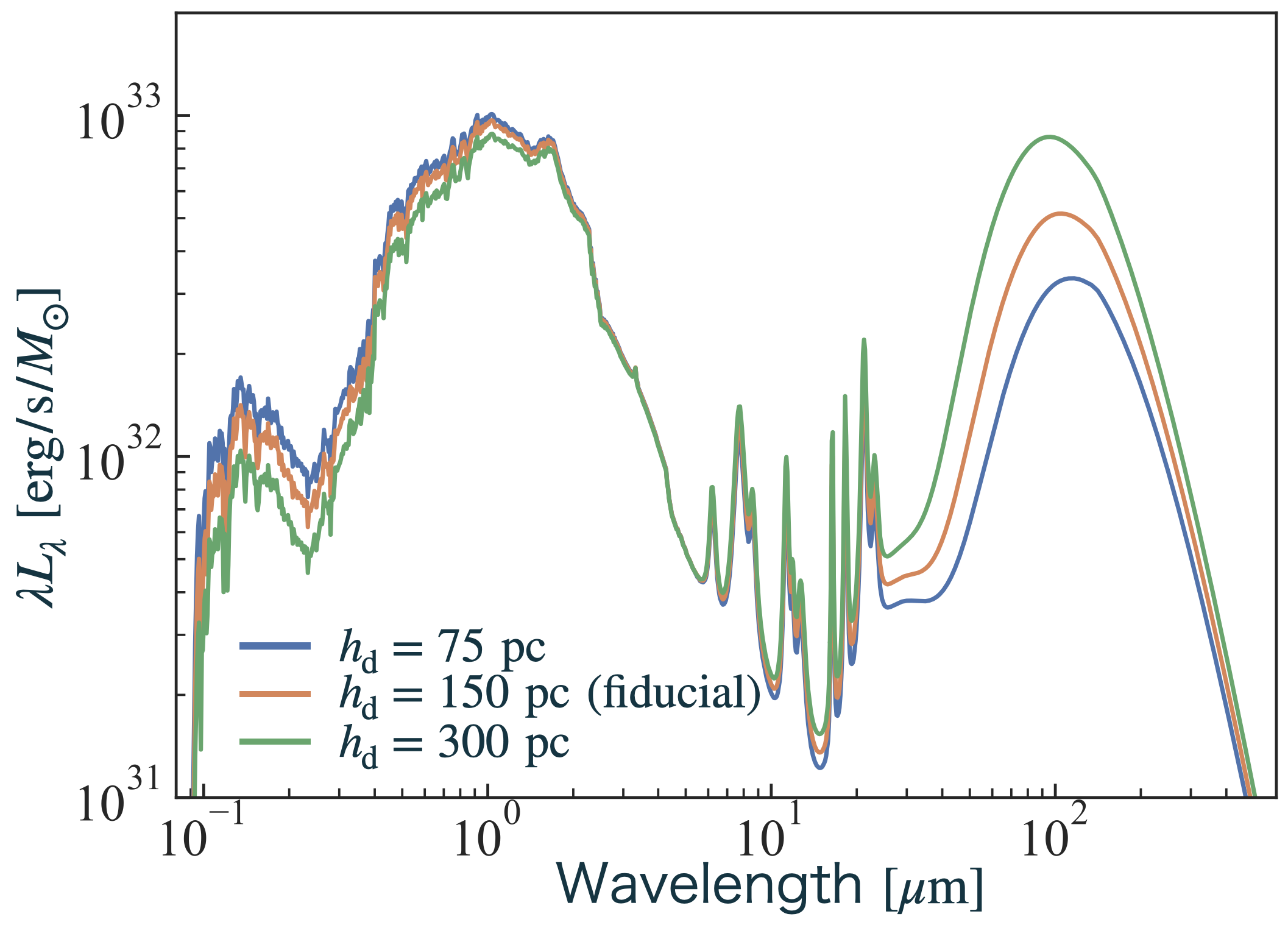}
    \caption{Galaxy SEDs with various dust scale heights $h_\mr{d}$ at the age of 13~Gyr.
    The star formation history is the same as Figure \ref{fig:SED_MW_at_13Gyr}.
    Blue, orange, and green curves represent 75, 150 (fiducial), and 300~pc, respectively. 
    }
	\label{fig:MW_h}
\end{figure}

Figure~\ref{fig:MW_h} shows the effect of changing the dust scale height of galaxy $h_\mr{d}$ for our galaxy SED model at an age of 13~Gyr.
The SFH is the same as Figure~\ref{fig:SED_MW_at_13Gyr}.
Blue, red, and green curves represent $h_\mr{d} =$ 75, 150 (fiducial), and 300~pc, respectively.
Intrinsic stellar radiation does not depend on $h_\mr{d}$.
Since the optical depth of the galaxy is defined as $\tau = \kappa_\mr{eff} h_\mr{d}$, $\tau$ increases as $h_\mr{d}$ increases.
Then, the absorption by the dust grain becomes stronger, and the observed UV radiation becomes weaker.
Since the energy absorbed by the dust grain increases, the radiation in the IR region becomes stronger.

Since our SED model assumes an axisymmetric one-dimensional disk, the galaxy has no structure in the radial direction and does not determine the radius.
However, in reality, when the radius of the galaxy changes and the volume changes, the density of the dust clump changes and the optical depth also changes.
In our model, the optical depth depends on clump filling fraction (Equation~(\ref{equ:f_cl})), and we assume that $n_\mr{H}$ is constant.
Therefore, if we consider a galaxy with a volume in which $n_\mr{H}$ changes significantly from 1~cm$^{-3}$, it is necessary to consider it, but it is not implemented in our model and is a future work.

\subsection{Effect of the ISM phase fraction}
\begin{figure}
	\includegraphics[width=\columnwidth]{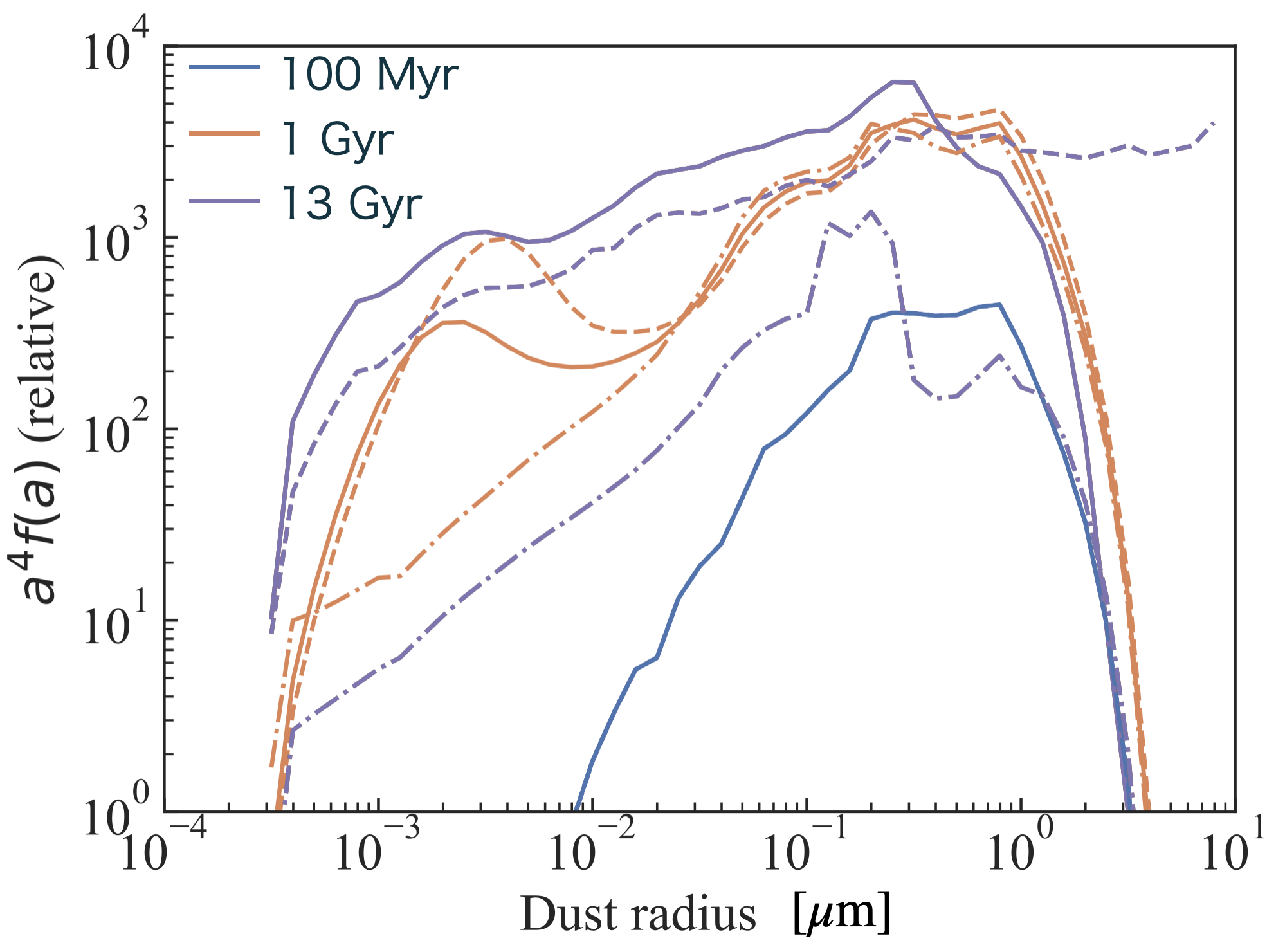}
	\caption{The dust size distribution with cold ISM region fraction $\eta = 0.5$ (fiducial, solid), 1 (dashed), and 0 (dot-dashed).
	Blue, orange, and purple curves are the age of 100~Myr, 1, and 13~Gyr galaxies, respectively.
	Note, the 100~Myr galaxy has three overlapping curves.
	}
\label{fig:effect_eta_size_distribution}
\end{figure}

\begin{figure}
	\includegraphics[width=\columnwidth]{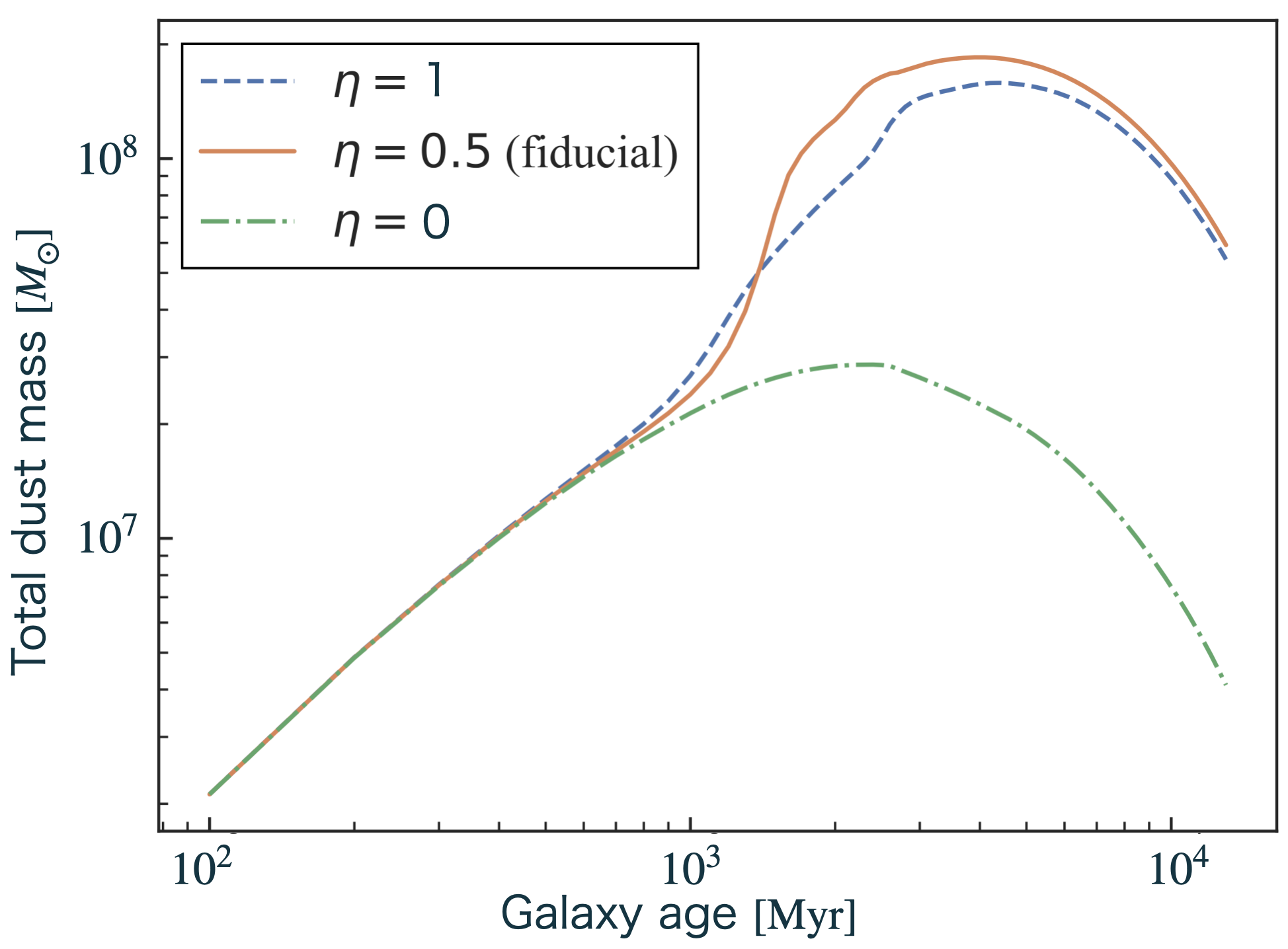}
	\caption{The evolution of total dust mass with $\eta = 0.5$ (fiducial, orange solid), 1 (blue dashed), and 0 (green dot-dashed).
	}
	\label{fig:effect_eta_total_dust_mass}
\end{figure}

In the current model, we consider three phases in the ISM: WNM, CNM and MC.
Figure \ref{fig:effect_eta_size_distribution} and \ref{fig:effect_eta_total_dust_mass} are the effects of the ISM phase fraction for dust size distribution.
The parameters except the ISM fraction are the same as the MW-like galaxy.
The value of the fraction of cold region $\eta$ is changed while keeping the ratio in the cold region constant to $\eta_\mr{CNM} : \eta_\mr{MC} = 3:2$ in this section.
Solid, dashed, and dot-dashed curves represent fiducial, $\eta = 1$ ($\eta_\mr{WNM} = 0.0$, $\eta_\mr{CNM} = 0.6$, and $\eta_\mr{MC} = 0.4$), and $\eta = 0$ ($\eta_\mr{WNM} = 1.0$, $\eta_\mr{CNM} = 0.0$, and $\eta_\mr{MC} = 0.0$) case, respectively.

At 100~Myr galaxy, there is no difference in three cases, since the stellar production dominates dust size distribution and is not affected by the ISM fractions.

At 1~Gyr galaxy, the size distribution of small $\eta$ cases have small amount of grain in radius of $> 2 \times 10^{-1}~\mr{\mu m}$ region because shattering is more likely to occur thanks to collisions between larger grains.
The bump in 10$^{-3}$--10$^{-2}$~$\mr{\mu m}$ is generated by accretion on the $> 10^{-3}$ size of the grain surface.
The bump is not observed in the $\eta = 0$ case, because in this case the accretion process on grains in the cold regions is not included.
The $\eta = 1$ result has a larger bump than the fiducial case.
This is because shattering is not effective yet, and the larger the fraction of cold regions is, the more effective the metal accretion.
On the contrary, large amounts of intermediate size grains ($2 \times 10^{-2}$--$2 \times 10^{-1}~\mr{\mu m}$) in a small $\eta$ case.
This results from the coagulation process in WNM.

At the 13~Gyr galaxy, a small amount of grain is observed in the $\eta = 0$ case.
When grain evolution occurs in only WNM, a strong shattering process generates the large amount of small grains.
Small dust grains are largely destroyed by SN shock \citep{Nozawa2006}, hence the mass of dust grains effectively decreases.
The large bump in $10^{-1}~\mr{\mu m}$ is caused by the balance between strong shattering and coagulation.
At the grain sizes of $< 1~\mr{\mu m}$, the size distribution of the $\eta = 1$ case has a smaller amount of dust.
This is because the shattering is weak in the $\eta = 1$ case and the metal accretion does not occur as effectively as the fiducial case, because the WNM is not considered in the calculation of the dust evolution.
Further, maximum grain radius in the $\eta = 1$ case reaches $> 1~\mr{\mu m}$, as weak shattering efficiency in CNM and MC.

The effect of $\eta$ for total dust mass of the MW-like galaxy model is shown in Figure \ref{fig:effect_eta_total_dust_mass}.
Very small amount of total dust mass is observed in the $\eta = 0$ case because the case does not consider the mass-increasing process other than the production from the stars.
Around 1~Gyr, the $\eta = 1$ case has a larger amount of total dust mass than the fiducial case.
This is because shattering is still less effective in this age, and metal accretion dominates the increase of total dust mass.
After 1~Gyr, the increase of total dust mass in the $\eta = 1$ case is slower than that in the fiducial case, since the rapid increase cycle is less effective in the $\eta = 1$ case.
The total mass at 13~Gyr is determined by the balance between destruction by SN shocks and the dust growth by metal accretion.

The galaxy SED at 13~Gyr with the cold ISM region fraction $\eta = 0.5$ (fiducial, solid orange), 1 (dashed blue), and 0 (dot-dashed green).
\begin{figure}
	\includegraphics[width=\columnwidth]{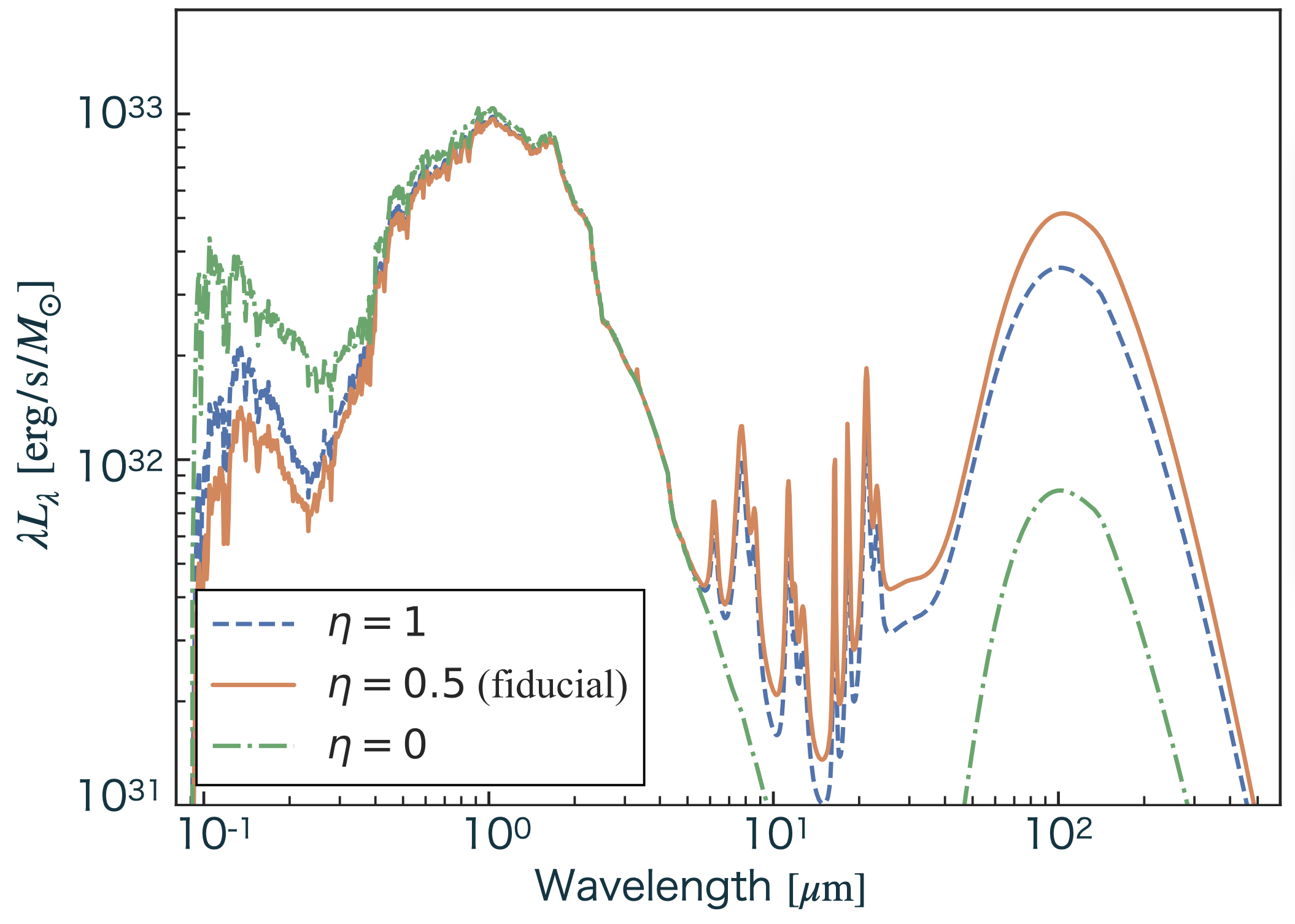}
	\caption{The galaxy SED at 13~Gyr with $\eta = 0.5$ (fiducial, orange solid), 1 (blue dashed), and 0 (green dot-dashed).
	The parameters are the same as the MW-like galaxy model except $\eta$.
	}
	\label{fig:effect_eta_SED}
\end{figure}
In $\eta = 0$ case, the SED has weaker dust attenuation and dust radiation than that of the fiducial case (Figure~\ref{fig:effect_eta_total_dust_mass}), since dust mass in all radius is smaller than the fiducial case.
Though the difference of total dust mass between the fiducial and the $\eta = 1$ case is small, the size distribution of the two cases has a large difference.
The $\eta = 1$ case has a lot of large dust grains and a few small dust grains.
The attenuation in the $0.1~\mr{\mu m}$ wavelength is mainly dominated by the $a < 1~\mr{\mu m}$ radius of grain.
As the large grain has large heat capacity, the radiation from the large grain is weaker than that from the small grain.
Therefore, $\eta = 1$ case has weaker attenuation in the UV region and weaker radiation in the IR region than the fiducial case.

\subsection{Effect of the coagulation threshold}
\label{sec:Effect of the coagulation threshold}

\begin{figure}
	\includegraphics[width=\columnwidth]{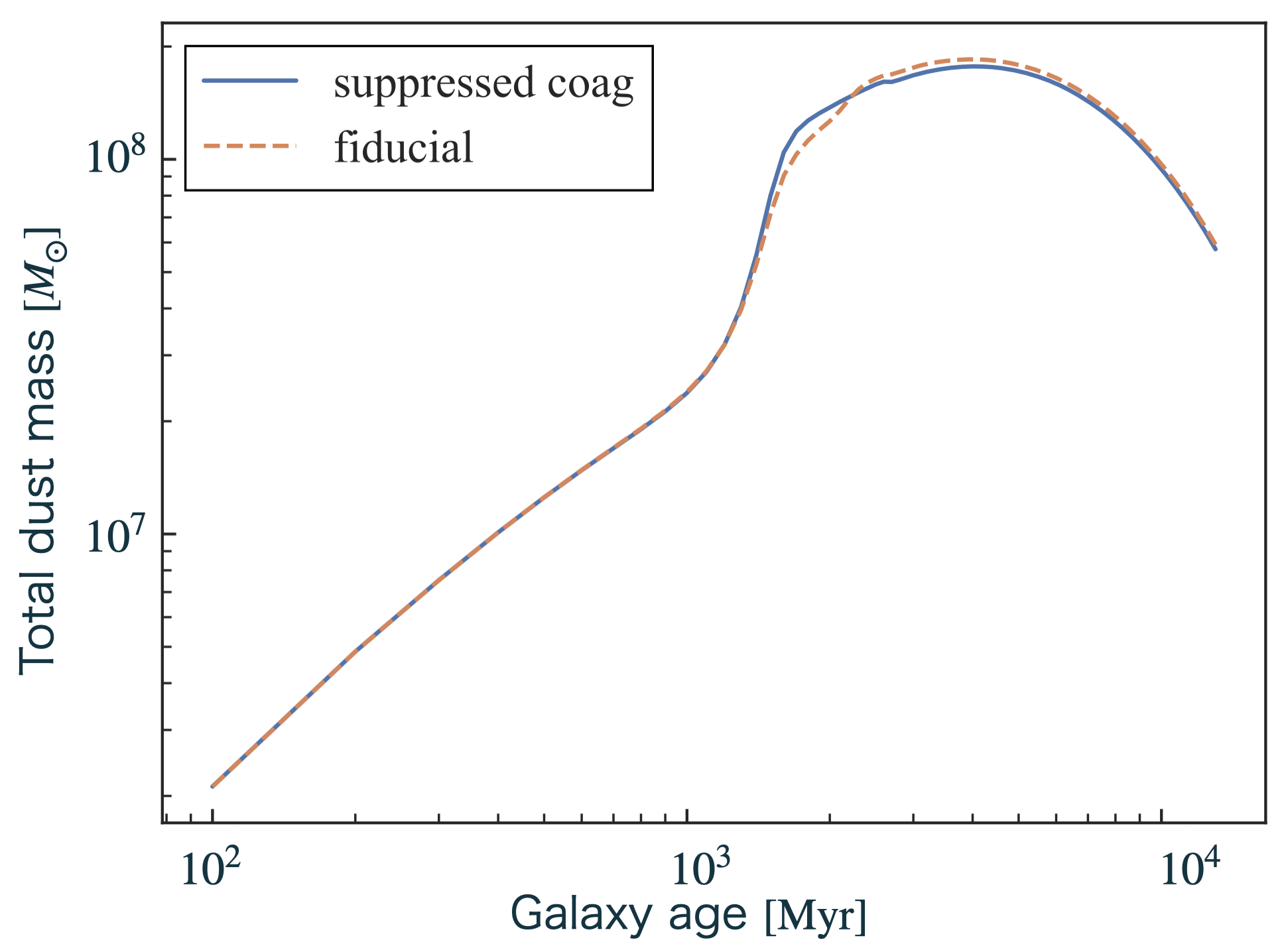}
	\caption{The effect of coagulation threshold velocity for total dust mass.
	Solid and dashed curves represent suppressed coagulation and fiducial case, respectively.
	The galaxy properties are the same as the MW-like galaxy model except $v_\mr{coag}$.
	}
\label{fig:mass_evolution_suppressed_coag}
\end{figure}

Coagulation can occur when the relative velocity of grains $v_\mr{coag}$ is slower than the threshold velocity.
However, we do not adopt the threshold of coagulation to reproduce the dust size distribution of the MW \citep{Asano2013a, Nozawa2015}.
If $v_\mr{coag}$ is adapted to the dust model, since the small radius grains have lower relative velocity, the small grains are more likely to coagulate.
Conversely, the large radius grains have large relative velocity and the grains cannot occur the coagulation.
\cite{Hirashita2009} show that the grain radius increases only up to 0.01--0.1~$\mr{\mu m}$, because the radii $a > 0.1~\mr{\mu m}$ grains have the relative velocity larger than $v_\mr{coag}$.
Therefore, a lower coagulation threshold velocity suppresses the effect of coagulation.
First, We show the effect of $v_\mr{coag}$ for total dust grain mass in Figure \ref{fig:mass_evolution_suppressed_coag}.
The dashed and solid lines represent the fiducial case (no $v_\mr{coag}$) and the adopted $v_\mr{coag}$ case (we call it a suppressed coagulation case).
Galaxy parameters are the same as \S\ref{sec:Result_of_dust_evolution_model} except $v_\mr{coag}$.
We calculate the coagulation velocity threshold in the same formula as \cite{Hirashita2009}.
$v_\mr{coag}$ between grain 1 and 2 is represented as 
\begin{equation}
    v_\mr{coag} = 21.4 \left[
    \frac{a^3_1 + a^3_2}{(a_1 + a_2)^3} \right]^{1/2}
    \frac{\gamma^{5/6}}{E^{1/3}R_{1,2}^{5/6}s^{1/2}},
\end{equation}
where suffix 1 and 2 represents the each value of grain 1 and 2, $R_{1,2} \equiv a_1a_2/(a_1 + a_2)$ is the reduced radius of the grains, $\gamma$ is the surface energy per unit area, and $E$ is related to the Poisson ratios ($\nu_1$ and $\nu_2$) and the Young modulus ($E_1$ and $E_2$) by $1/E \equiv (1-\nu_1)^2/E_1 + (1 - \nu_2)^2/E_2$.
The value of $\gamma$, $\nu$, and $E$ are 25~$\mr{erg/cm^{2}}$, 0.17~$\mr{erg/cm^{2}}$ and $5.4 \times 10^{11}~\mr{dyn/cm^{2}}$ for silicate, and 12~$\mr{erg/cm^{2}}$, 0.5 and $3.4 \times 10^{10}~\mr{dyn/cm^{2}}$ for graphite from \cite{Chokshi1993}.
From Figure~\ref{fig:mass_evolution_suppressed_coag}, the effect of $v_\mr{coag}$ for total dust mass is very small.
Coagulation itself decrease total surface area of grains and suppress the cross section of the metal accretion.
On the other hand, as the grain size increases, shattering is more likely to occur, and smaller radius grain increases.
Since these effects are balanced, coagulation only suppresses the increase in total dust mass and has a small effect.
Coagulation is difficult to occur in young galaxies (age > 1~Gyr), and becomes effective after the shattering process becomes effective.
Therefore, when the coagulation becomes effective, the rapid increase in the total dust grain mass has already finished, and the coagulation does not significantly affect the total mass, but only changes the size distribution of the dust grain.

\begin{figure}
	\includegraphics[width=\columnwidth]{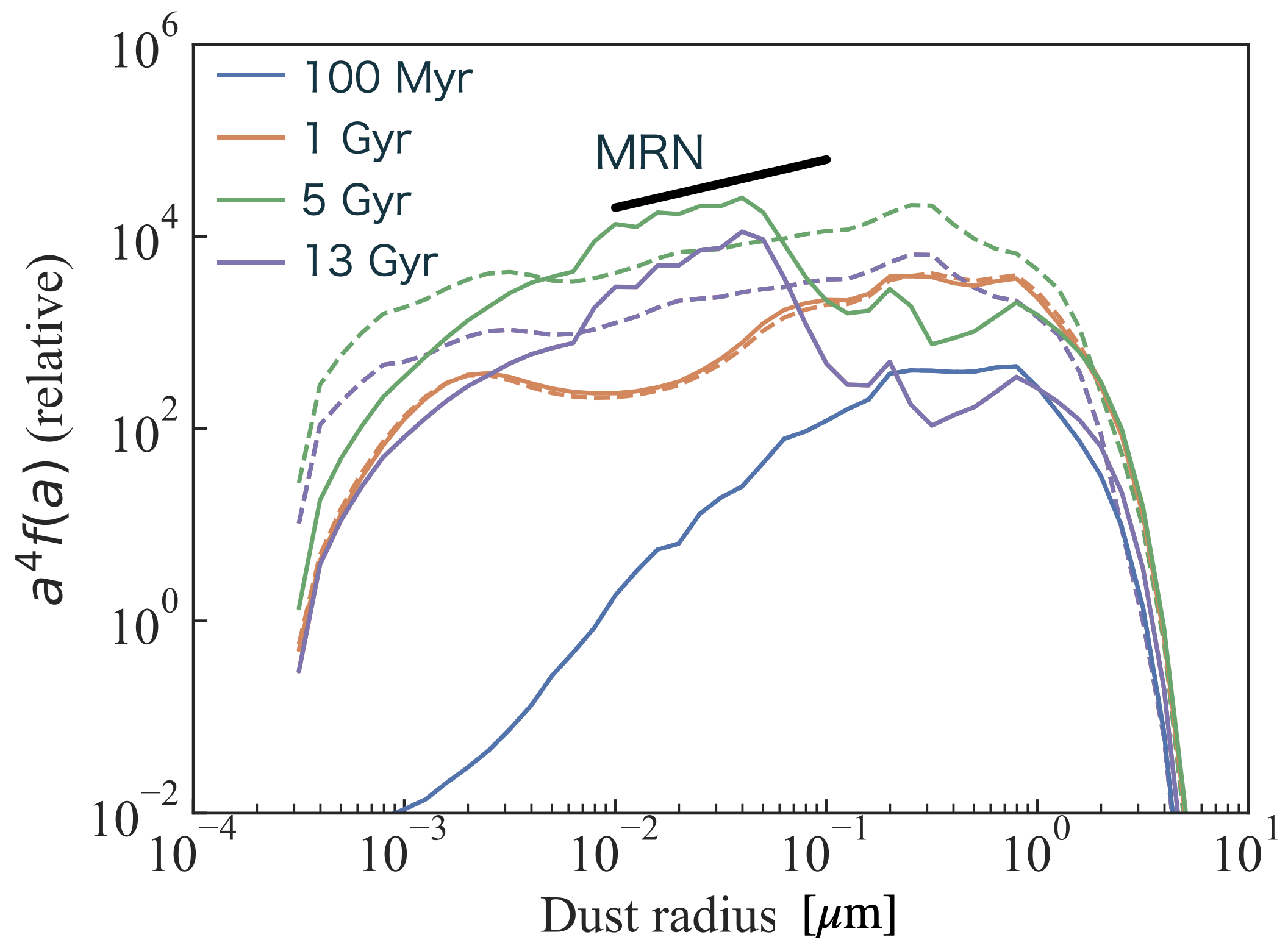}
	\caption{The effect of $v_\mr{coag}$ for dust size distribution.
    Blue, orange, green, and purple curves represent the dust grain size distribution with the age of 100~Myr, 1~Gyr, 5~Gyr, and 13~Gyr, respectively.
    Solid and dashed lines represent suppressed coagulation and fiducial case, respectively.
    }
\label{fig:dust_distribution_suppressed_coag}
\end{figure}

Second, we show the effect of $v_\mr{coag}$ for dust size distribution in Figure~\ref{fig:dust_distribution_suppressed_coag}.
In the age of the 100~Myr and 1~Gyr galaxies, since effective dust evolution in the ISM has not started yet, there is no difference in the dust size distribution between the two cases.
On the other hand, $v_\mr{coag}$ strongly affects the dust distribution after 1~Gyr galaxy.
$v_\mr{coag}$ suppresses the coagulation between larger grains and determines the maximum radius of dust grain.
Coagulation shifts the size distribution to larger sizes, thus, in the suppressed coagulation case, the dust size distribution is biased toward the smaller one, and the slope is also different from MRN.
Therefore, adopting the low $v_\mr{coag}$ in the MW-like galaxy model leads to the dust size distribution to be different from the MRN distribution, and thus we do not adopt $v_\mr{coag}$ in our model.

\subsection{Comparison between closed-box and infall model}
\label{sec:Comparison_between_closed-box_and_infall_model}

\begin{figure}
    \includegraphics[width=\columnwidth]{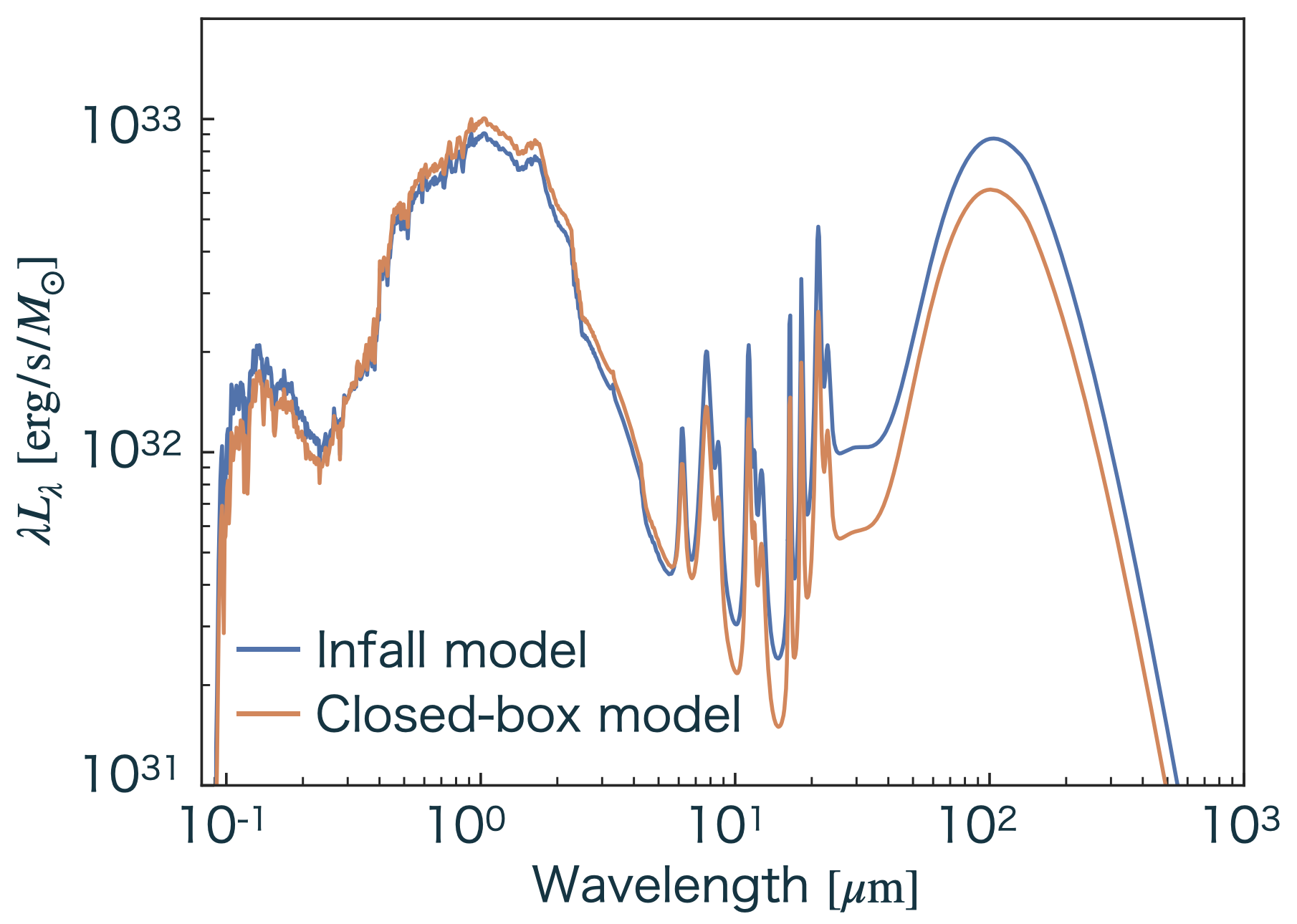}
    \caption{The comparison of MW-like model galaxy SED at age $t_\mr{gal} = 13$~Gyr with infall and closed-box model. 
    The orange curve represents closed-box model (same as Figure \ref{fig:SED_MW_at_13Gyr}), the blue curve represents infall model with infall time scale $\tau_\mr{infall} = 6$~Gyr.}
    \label{fig:SED_MW_comparison}
\end{figure}

\begin{figure}
    \includegraphics[width=\columnwidth]{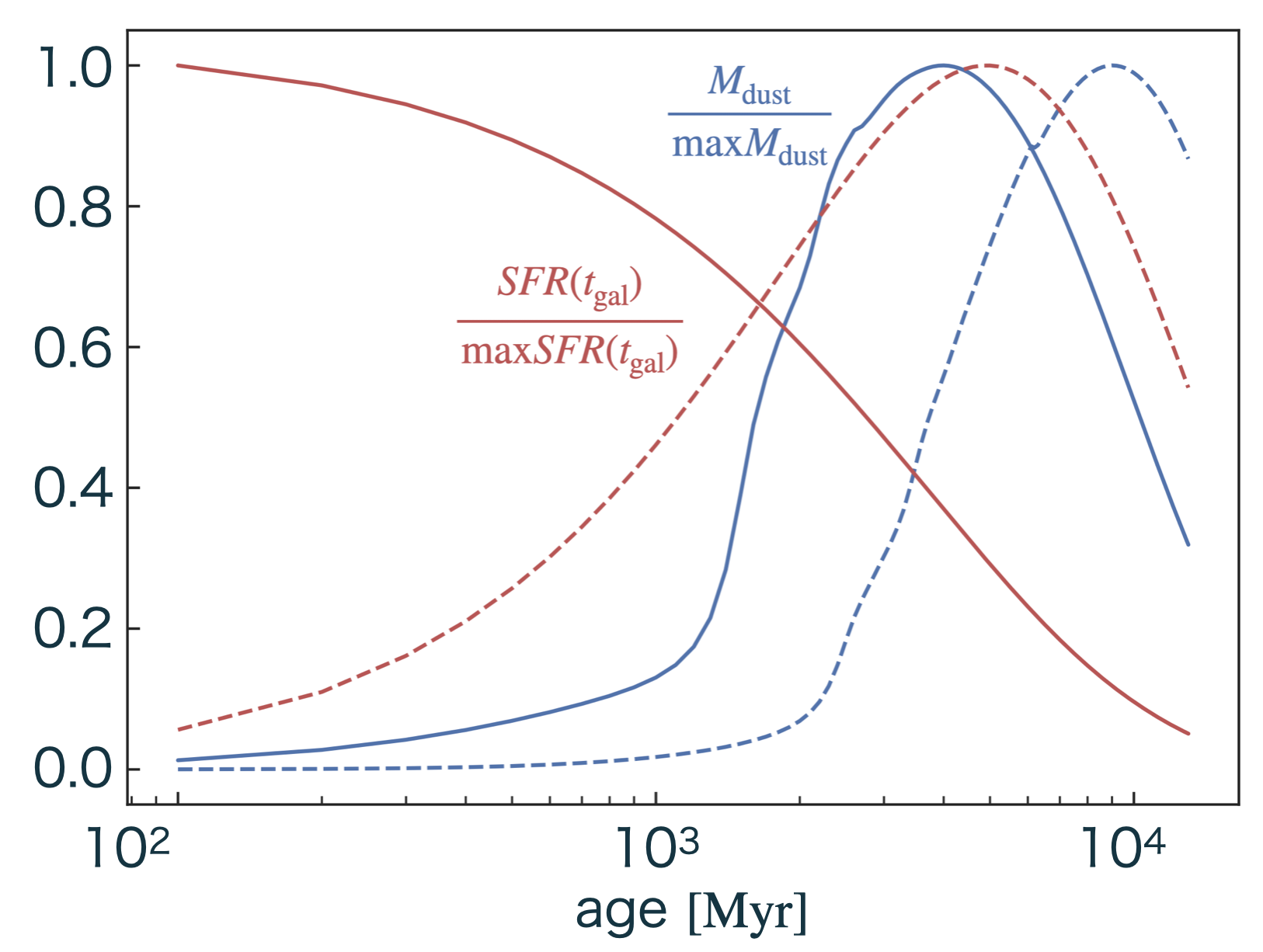}
    \caption{The time evolution of dust mass and SFR of the galaxy with closed-box and infall model (same galaxies as Figure \ref{fig:SED_MW_comparison}).
    Colors represent the difference of quantities: the ratio of $\mr{SFR}(t_\mr{gal})$ and maximum value of it (red), and the ratio of dust mass $M_\mr{dust}$ and maximum value of it (blue).
    The solid and dashed curves represent closed-box and infall models, respectively.}
    \label{fig:comparison_close_open_mass}
\end{figure}

The comparison between the result of the closed-box and infall model is shown in Figure~\ref{fig:SED_MW_comparison}.
We adopt the following equation for the infall rate \citep{Inoue2011}:
\begin{equation}
    \diff{M_\mr{infall}}{t} = \frac{M_\mr{infall}}{\tau_\mr{infall}}
    \exp\left( -\frac{t}{\tau_\mr{infall}}\right), 
\end{equation}
where $\tau_\mr{infall}$ is the timescale of infall, and $M_\mr{infall}$ is the total mass that flows into the galaxy by infall as $t \rightarrow \infty$.
For the infall model, the initial mass of a galaxy is set to zero, and primordial gas (zero-metallicity) fall onto the galaxy with $M_\mr{infall} = 10^{11}~M_\odot$ and $\tau_\mr{infall} = 6$~Gyr.
The time evolution of the SFR and dust mass are plotted in Figure~\ref{fig:comparison_close_open_mass}.
The star formation history is very different between the two models.
While the SFR of the closed-box model case monotonically decrease, the SFR of the infall model gradually increase and it reaches to a peak at $t_\mr{gal} = 5$~Gyr (close to infall timescale $t_\mr{infall} = 6$~Gyr), after that the SFR decrease gradually.
Since the SFR at 13~Gyr of the infall model is higher than that of the closed-box model, the ratio of younger stars is increased in the infall model, and then the luminosity of the UV region becomes stronger.
On the other hand, the continuum at near IR wavelengths emitted from old stars becomes slightly weaker due to the smaller amount of old stars.
The metallicity is 1.6~$Z_\odot$ in the closed box model, while it is 0.86~$Z_\odot$ in the infall model which is closer to the solar metallicity.

In the infall model case, the peak of the dust mass comes later because of the different star formation history.
It leads to an increase of the IR emission which is emitted by dust grains.
In general, the infall model tends to delay the evolution of the galaxy.

\subsection{Radio emission}

Our model does not include the radio emission.
Because, in a normal galaxy, the luminosity of the radio region is only $< 10^{-4}$ of the overall bolometric luminosity of a galaxy \citep{Condon1992}.
Above $\sim 1$~mm, radio emission is swamped in dust emission for normal galaxies.
The radio is mainly emitted by synchrotron radiation from relativistic electrons accelerated in supernova remnants and free-free emission by H{\sc{ii}} region, which is a ionized by the radiation of heavy and young stars \citep{Klein1988, Carlstrom1991}.
Since both radio sources are associated with SN explosions, their radiation is considered to depend on the SN rate (SNR) \citep{Condon1992}.
In particular, many galaxies with strong synchrotron radiation by a jet from active galactic nuclei have been observed \citep[e.g.,][]{Carilli1991, Laing2002}, and we will take it into account in our future work.

\section{Conclusions}
In this paper, we construct a new galaxy SED model including the dust evolution in galaxies consistent with the chemical evolution \citep{Asano2013a, Asano2013b, Asano2014}.
The dust model considers several evolutionary processes of the dust production by AGB stars and SNe II, the destruction by SN shocks in the ISM, the grain growth by metal accretion to grain surface, and the two types of grain-grain collision, the shattering and coagulation.
The stellar radiation is calculated by P\'EGASE.2~\citep{Fioc1999}.
Based on this, we constructed a radiative transfer model with a one-dimensional plain parallel geometry equipped with the mega-grain approximation for fast computation \citep{Varosi1999, Inoue2005}.
For the radiation from dust, we take into account the stochastic heating of dust grains by Monte Carlo simulation.
As a fiducial model, we assumed the Schmidt law with star formation time scale $\tau_\mr{SF} = 3~\mr{Gyr}$, the Salpeter IMF \citep{Salpeter1955}, and the closed box model.
The ISM phase fractions were set as $\eta_\mr{WNM} = 0.5$, $\eta_\mr{CNM} = 0.3$, and $\eta_\mr{MC} = 0.2$, scale height of dust is $h_\mr{d} = 150$~pc, and the threshold of coagulation velocity is removed.
Our model indicates that early galaxies ($\sim 100$~Myr) produce a small amount of dust.
The PAHs, which dominate the MIR wavelength region, have not been produced yet, in particular.
The SED at the age of 100~Myr is dominated by stellar emission.
Then the amount of dust mass and emission explosively increases at the age of about 3~Gyrs.
Subsequently, the dust mass and emission from both the stars and dust decreases, along with the decline of the star formation rate.
Since this model treats the evolution of dust appropriately, we can apply it to any age of a galaxy as far as the model assumptions are valid.

\section*{Acknowledgements}
First of all, we offer our sincere thanks to the anonymous referee for her/his enormous effort to read through the article and invaluably important comments and suggestions that improved the quality of the paper very much.  
We are grateful to the colleagues in the Lab for fruitful discussions and comments.
We thank H. Kobayashi and A.K. Inoue for helpful comments on the coding of dust evolution model.
This work has been supported by JSPS Grants-in-Aid for Scientific Research (17H01110, 19H05076, and 21H01128). 
This work has also been supported in part by the Sumitomo Foundation Fiscal 2018 Grant for Basic Science Research Projects (180923), and the Collaboration Funding of the Institute of Statistical Mathematics ``New Development of the Studies on Galaxy Evolution with a Method of Data Science''.

\section*{Data Availability}
The data underlying this article will be shared on reasonable request to the corresponding author.
 



\bibliographystyle{mnras}
\bibliography{library} 








\bsp	
\label{lastpage}
\end{document}